\documentclass[letterpaper]{aa}
\usepackage{graphicx,txfonts}

\begin{document}

\title{(An)isotropy of the Hubble diagram: comparing hemispheres}

\author{
   Dominik J.~Schwarz\thanks{\tt dschwarz at physik dot uni-bielefeld dot de}
   \and
   Bastian Weinhorst\thanks{\tt bwein at physik dot uni-bielefeld dot de}}

\institute{
Fakult\"at f\"ur Physik, Postfach 100131, Universit\"at Bielefeld,
33501 Bielefeld, Germany}

\date{Received ddmmyy / Accepted ddmmyy}

\abstract
{}
{We test the isotropy of the Hubble diagram. At small redshifts, this is
possible without assumptions on the cosmic inventory and provides a
fundamental test of the cosmological principle. At higher redshift we
check for the self-consistency of the $\Lambda$CDM model.}
{At small redshifts, we use public supernovae (SNe) Ia data to determine the
deceleration parameter $q_0$ and the SN calibration on opposite
hemispheres. For the complete data sets we fit $\Omega_{\rm M}$ and the
SN calibration on opposite hemispheres.}
{A statistically significant anisotropy of the Hubble diagram at redshifts
$z < 0.2$ is discovered  ($> 95\%$C.L.). While data from the North Galactic
hemisphere favour the accelerated expansion of the Universe, data from the
South Galactic hemisphere are not conclusive. The hemispheric asymmetry is
maximal toward a
direction close to the equatorial poles. The discrepancy between the
equatorial North and South hemispheres shows up in the SN calibration. For the
$\Lambda$CDM model fitted to all available SNe, we find the same asymmetry.}
{The alignment of discrepancies between hemispheric Hubble diagrams with the
equatorial frame seems to point toward a systematic error in the SN search,
observation, analysis or data reduction. We also find that our model
independent test cannot exclude the case of the
deceleration of the expansion at a statistically significant level.}


\maketitle

\section{Introduction}

The Hubble law is a direct consequence of the cosmological principle.
Modern Hubble diagrams from supernovae Ia (SNe Ia) confirm the Hubble
law and provide evidence for an {\em accelerated} expansion of the Universe
(Riess et al.~\cite{Riess:1998cb}; Perlmutter et al.~\cite{Perlmutter:1998np}).
In these studies the isotropy of the Hubble diagram is assumed. The purpose
of this work is to provide quantitative tests of the isotropy of SNe Ia
Hubble diagrams, beyond the identification of the cosmic microwave dipole in
the local SNe Ia data (Riess et al. \cite{Riess:1995}).

A test of the isotropy of Hubble diagrams is interesting for at
least three reasons: 1.~for checking the validity of the
cosmological principle (Kolatt \& Lahav~\cite{Kolatt}),
2.~to measure expected deviations from the
isotropy [due to Local Group motion, peculiar velocities, and other
effects from structure formation (Sasaki \cite{Sasaki:1987};
Radburn-Smith et al.~\cite{Radburn}; Bonvin at al.~\cite{Bonvin:2005ps};
Hui \& Greene~\cite{Hui}; Weinhorst~\cite{weinhorst};
Cooray \& Caldwell~\cite{Cooray}; Haugb\o lle
et al.~\cite{haugbolle}; Neill et al.~\cite{Neill}; Wang~\cite{Wang};
Hannestad et al.~\cite{Hannestad}, Gordon et al.~\cite{Gordon})],
3.~to search for systematic errors in the observations and their
analysis (Kolatt \& Lahav~\cite{Kolatt}, Gupta et al.~\cite{Gupta}).

The cosmological principle states that the statistical distribution
of matter and light in the Universe is isotropic and homogeneous in space.
The principle itself may either be motivated by the idea of cosmological
inflation or by a simplicity argument.
As a consequence of the ergodic theorem (spatial averaging replaces ensemble
averaging) the spatial isotropy and homogeneity of the statistical
distributions becomes an approximate symmetry of the space-time metric
and matter distribution at large scales.

Consequently, the cosmological principle implies that distances, angles,
time intervals etc.~at large scales can be measured
according to the ruler sticks and clocks described by the Robertson-Walker
line element.  The redshift of light from distant (with the expansion comoving)
objects and the Hubble law for redshift $z \ll 1$ follow directly
(without making use of Einstein's equation), i.e.~the Hubble law does not
depend on the details of the cosmological model (like the inventory of the
Universe).  The Hubble diagram (distance or magnitude versus redshift)
therefore provides one of the most fundamental tests of modern cosmology.
It provides a test of the cosmological principle and the idea that the
space-time is correctly modelled as a riemannian manifold.

However, the local Universe is neither isotropic nor homogeneous.
Consequently, the observations are expected to approach the Hubble law
at some $z>0$ only.  Deviations from the isotropic and homogeneous
Universe arise from structure formation and can be described by linear
perturbation theory at large scales.  Thus in order to measure the Hubble
constant, the deceleration parameter and other cosmological parameters
of an isotropic and homogenous model, it is {\em compulsory} to demonstrate
that the effects from structure formation are either negligible or
properly considered in the error bars.

Most importantly, one would like to firmly establish the acceleration of the
Universe in a model independent way; i.e.~to measure the kinematics of the
Universe without assuming the validity of Einstein's equation and without
assumptions on the matter content of the Universe. This leads to a dilemma:
at very small redshifts, the Hubble law does not hold, since the local
Universe is inhomogeneous. At high redshift, $z \sim 1$, the luminosity
distance $d_{\rm L}$ (or equivalently the distance modulus $m - M$)
depends on the detailed cosmological model. Therefore, such an analysis
must be restricted to a finite range of redshifts below one, but must exclude the most local
SNe.

A simple and powerful method to establish the existence of such a range of
redshifts is to test the isotropy of the Hubble diagram.
To be more precise, the isotropy of the
Hubble diagram is a necessary, but not a sufficient condition for the
existence of such an interval. Here we use SNe Ia data to test the isotropy
of the Hubble diagram on opposite hemispheres of the sky.

This test is also closely related to the issue of fairly sampling the Universe
with SNe. In other words, is it suitable to determine the global Hubble
rate from a local measurement.  The 1998 cosmology-revolution
(Riess et al.~\cite{Riess:1998cb}; Perlmutter  et al.~\cite{Perlmutter:1998np})
relies on the assumption that
the SNe Ia at low redshifts (say $z<0.2$) represent a fair sample of the
Universe and that the Hubble law is a good approximation to the data.
Based on SNe Ia at low and high redshifts together, it has been
concluded that the present expansion of the Universe accelerates.

At the same time, our test provides a cross check for the measurements of the
Hubble constant $H_0$ and the deceleration parameter $q_0$.  The most
commonly adopted values for the inflationary $\Lambda$ cold dark matter
model ($\Lambda$CDM) are from the HST key project $H_0 = 72 \pm 8$ km/s/Mpc
(Freedman et al.~\cite{Freedman:2000cf}) and from a fit to WMAP data
$H_0 = 73^{+3}_{-4}$ km/s/Mpc and $q_0 = (3\Omega_{\rm M} - 2)/2 =
-0.64^{+0.04}_{-0.06}$ (derived, not directly measured) (Spergel et
al.~\cite{Spergel:2006hy}).
The calibration of SNe Ia used in the HST key project analysis
has recently been criticised by Sandage et al.~\cite{Sandage:2006cv}.
The HST key project analysis is based on a set of cepheides from a
metal-poor environment, whereas most of the SNe of the sample are in
metal-rich galaxies. Sandage et al.~\cite{Sandage:2006cv} obtain
$H_0 = 62.3 \pm 1.5 \pm 5.0$ km/s/Mpc, using cepheides from a metal-rich
environment and a significantly larger number of calibrators and SNe.
At face value, there is now some tension between CMB measurements and the
direct determination from the Hubble diagram. A possible explanation would
be that the local value does not coincide with the global one. Our test
is suited to detect such a situation, as we would expect it to go along with an
anisotropy of the Hubble diagram.

In the second part of this work we present a model dependent analysis of
the isotropy of Hubble diagrams, relying on the spatially flat $\Lambda$CDM
model, i.e. a model with two free parameters $H_0$ and $\Omega_{\rm M}$.

The paper is structured as follows. In section \ref{Hubble} we comment
on some basic properties of the Hubble law.
In section \ref{data} we describe the four SNe Ia data sets that we are using. The
method of our tests is explained in section \ref{method}
and section \ref{results} contains the results. We conclude in section
\ref{conclusion}.

\section{Hubble law}
\label{Hubble}

The luminosity distance $d_{\rm L}$ is a function of redshift and angular
position of an observed object.  It is linked to the distance
modulus $m - M = 5 \log(d_{\rm L}/1 {\rm Mpc}) + 25$. The observed
redshift $z$ of the object contains information on its peculiar velocity,
the peculiar velocity of the observer and the cosmological redshift.
If we assume that the
cosmological principle holds and that the observed objects are comoving
with the expansion of the Universe, one finds the Hubble law
\begin{eqnarray}
d_{\rm L}(z) &=& \frac{c}{H_0} \left[z + \left(1-q_0\right) \frac{z^2}{2} +
\right. \nonumber \\
&& \left. \left(-j_0 +3q_0^2 +q_0 -1 -\frac{k}{a_0^2} \frac{c^2}{H_0^2}\right)
\frac{z^3}{6} + {\cal O}(z^4)\right],
\label{HubbleLaw}
\end{eqnarray}
where $j$ denotes the jerk, $k = 0,\pm 1$ the normalised,
comoving curvature and $a$ the scale factor. In terms of the scale
factor and its derivatives
\begin{equation}
H = \frac{\dot{a}}{a}, \quad q = - \frac{\ddot{a}}{a} \frac{1}{H^2}, \quad
j = \frac{\dot{\ddot{a}}}{a}\frac{1}{H^3}.
\end{equation}
We see that the luminosity distance does not depend on
the geometry of the Universe up to order $z^2$, while
the terms of order $z^3$ depend on jerk and geometry. We restrict our
analysis to a test of the Hubble law up to order $z^2$.
A useful reference is the empty-Universe model, which is characterised
by $q=j=0$. Using the Einstein equation for the reference model
we find additionally $H^2/c^2 = -k/a^2$, thus $k=-1$ and
$H_0 d_{\rm L}(z) = cz(1+z/2)$.

Let us estimate the error from neglecting jerk and curvature in
(\ref{HubbleLaw}). For a general $\Lambda$CDM model, we find
$q_0 = (\Omega_{\rm M} - 2 \Omega_\Lambda)/2$ and $j_0 =
\Omega_{\rm M} + \Omega_\Lambda$. In the special case
of the Einstein-de Sitter model ($q_0 = 1/2, j_0 = 1, k=0$), the ratio of
the third order term to the second order term in (\ref{HubbleLaw})
becomes $-1/2$ and thus the error from neglecting the third order term amounts
to $10\%$ at $z = 0.2$. For the WMAP best-fit flat $\Lambda$CDM model
the error at the same redshift is $6\%$. Therefore, in order
to obtain $q_0$ and its confidence contours at $10\%$ theoretical
accuracy or better, we must restrict the {\it model independent}
fits to SNe at redshifts below $z=0.2$.

For the spatially flat $\Lambda$CDM model (assuming Einstein's equation)
the exact expression for the luminosity distance is given by
\begin{equation}
\label{Hubble2}
d_{\rm L}(z) = c (1+z) \int_0^z \frac{{\rm d}z'}{H(z')},
\end{equation}
with $H(z)^2 = H_0^2 [\Omega_{\rm M} (1+z)^3 + 1 - \Omega_{\rm M}]$. Thus
this model is fully characterised by the parameters $H_0$ and
$\Omega_{\rm M}$, which can be easily compared to the model independent
parameters $H_0$ and $q_0$. Expression (\ref{Hubble2}) is the basis
of our {\it model dependent} fit and, in contrast to (\ref{HubbleLaw}),
holds for arbitrary redshift.

The effect of peculiar velocities on the redshift can be incorporated
by realising that
\begin{equation}
c(1+ z_{\rm obs}) =  c (1+z_{\rm com})\left[1 +
{\bf \hat e} ({\bf v}_{\rm pec} - {\bf v}_\odot)/c + \dots \right] ,
\end{equation}
where ${\bf \hat e}$ denotes the direction of observation (unit vector)
--- light propagates along $- {\bf \hat e}$ --- and the dots stand
for higher order terms in the peculiar velocities.
As shown by Bonvin et al.~(\cite{Bonvin:2005ps,Bonvin:2006en}),
it is expected that the peculiar motion of the Solar system w.r.t.~the CMB
is the dominant source of anisotropy in the Hubble diagram. Indeed the
CMB dipole has its correspondence in SN Ia data
(Riess et al.~\cite{Riess:1995}). It is easily
subtracted, assuming that the CMB dipole is entirely due to the Sun's motion
w.r.t.~the CMB. We take the WMAP 1yr value for the dipole:
$(l,b) = (263.85^\circ \pm 0.1^\circ, 48.25^\circ \pm 0.04^\circ)$ and
$v_\odot/c = T_1/T_0 = [(3.346 \pm 0.017)/(2.725\pm 0.001)] =
1.23 \times 10^{-3}$ (Bennett et al.~\cite{Bennett:2003bz}). At small
redshifts it is sufficient to work with
\begin{equation}
c z_{\rm cmb} =  c z_{\rm com} + {\bf \hat e} {\bf v}_{\rm pec} + \dots
\end{equation}
We assume that the peculiar velocities are uncorrelated with zero mean
and variance $\sigma_v^2$.

\begin{figure}
\centering
\includegraphics[width=0.8\linewidth]{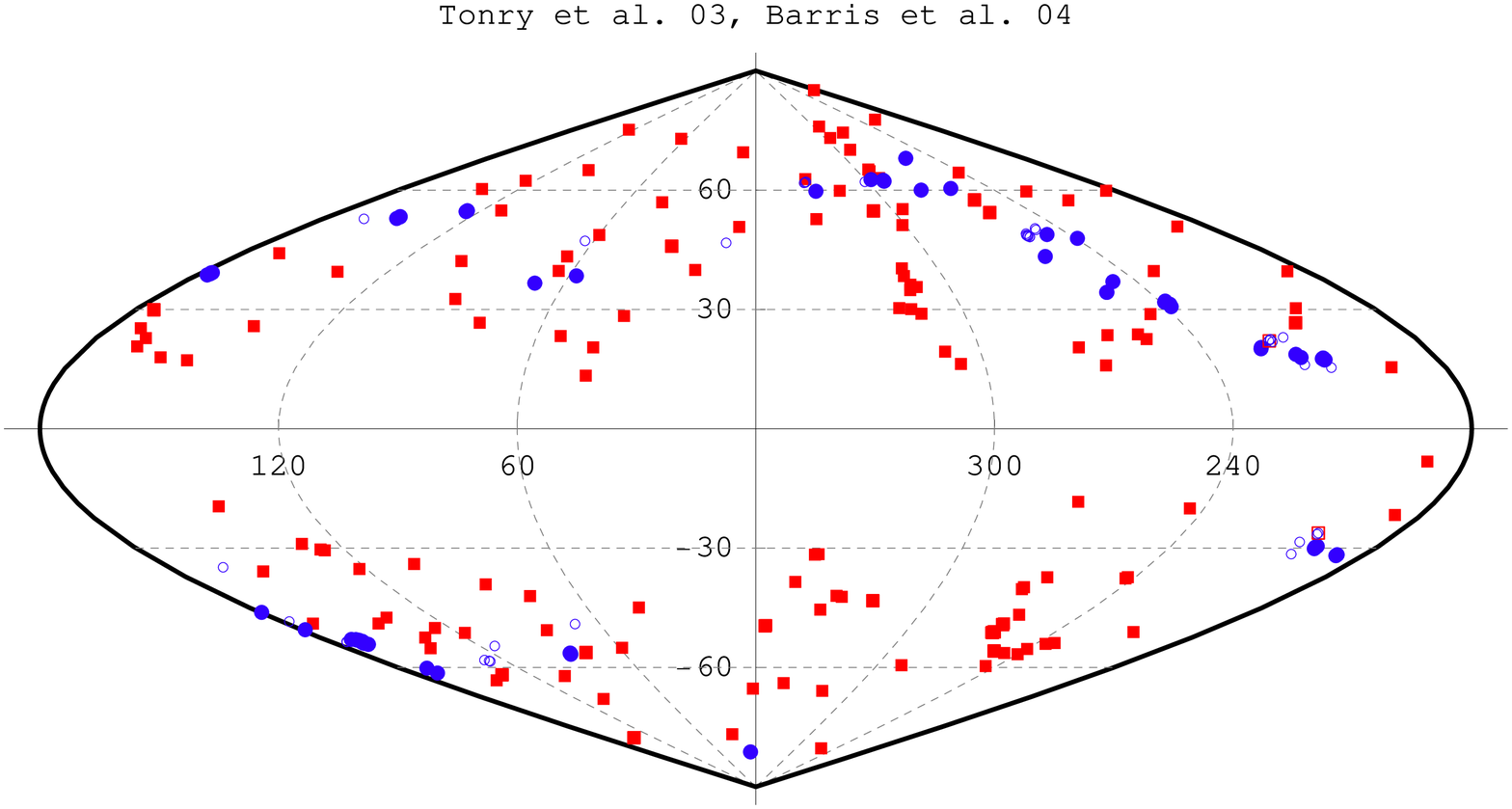}
\includegraphics[width=0.8\linewidth]{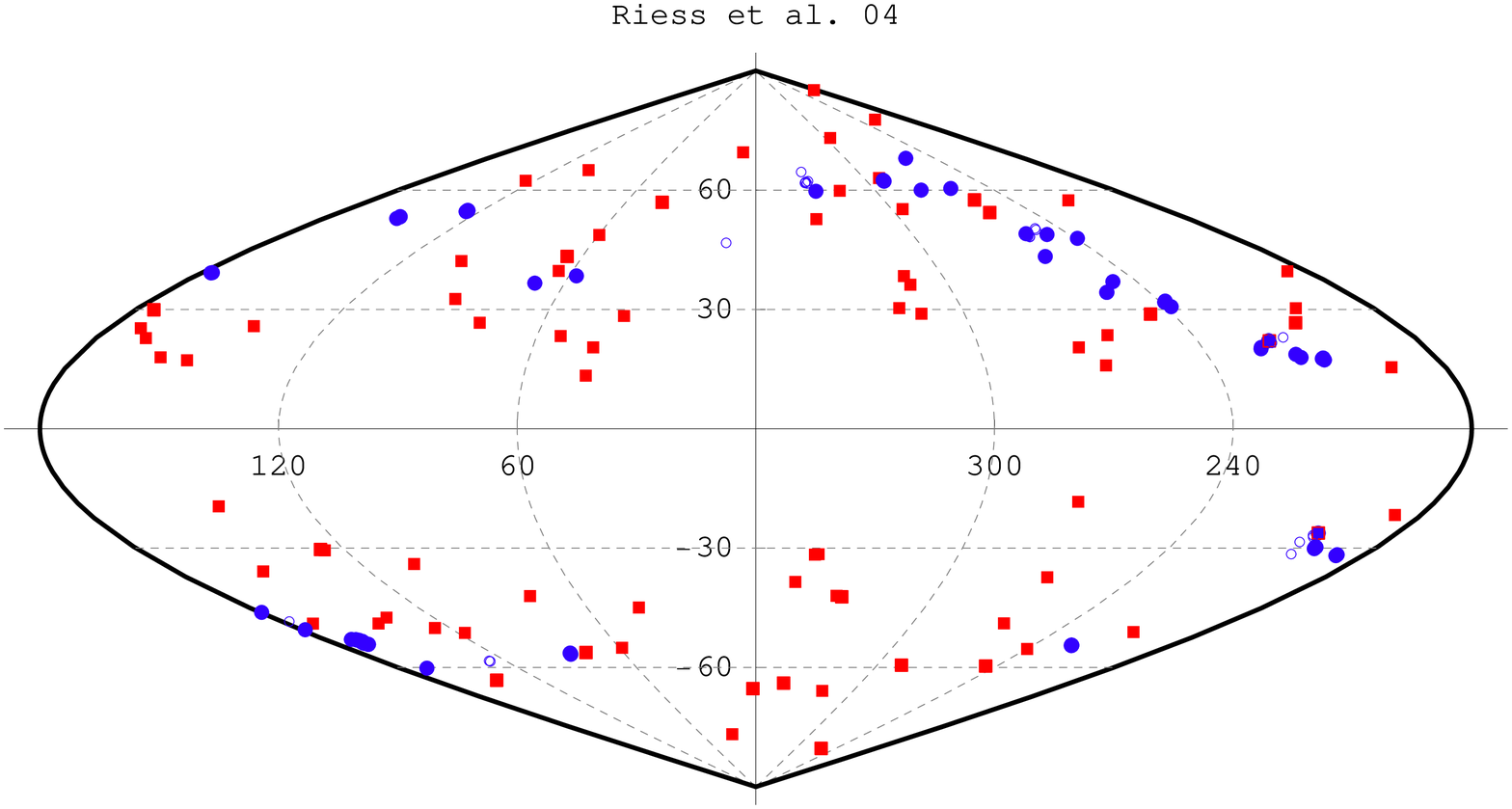}
\includegraphics[width=0.8\linewidth]{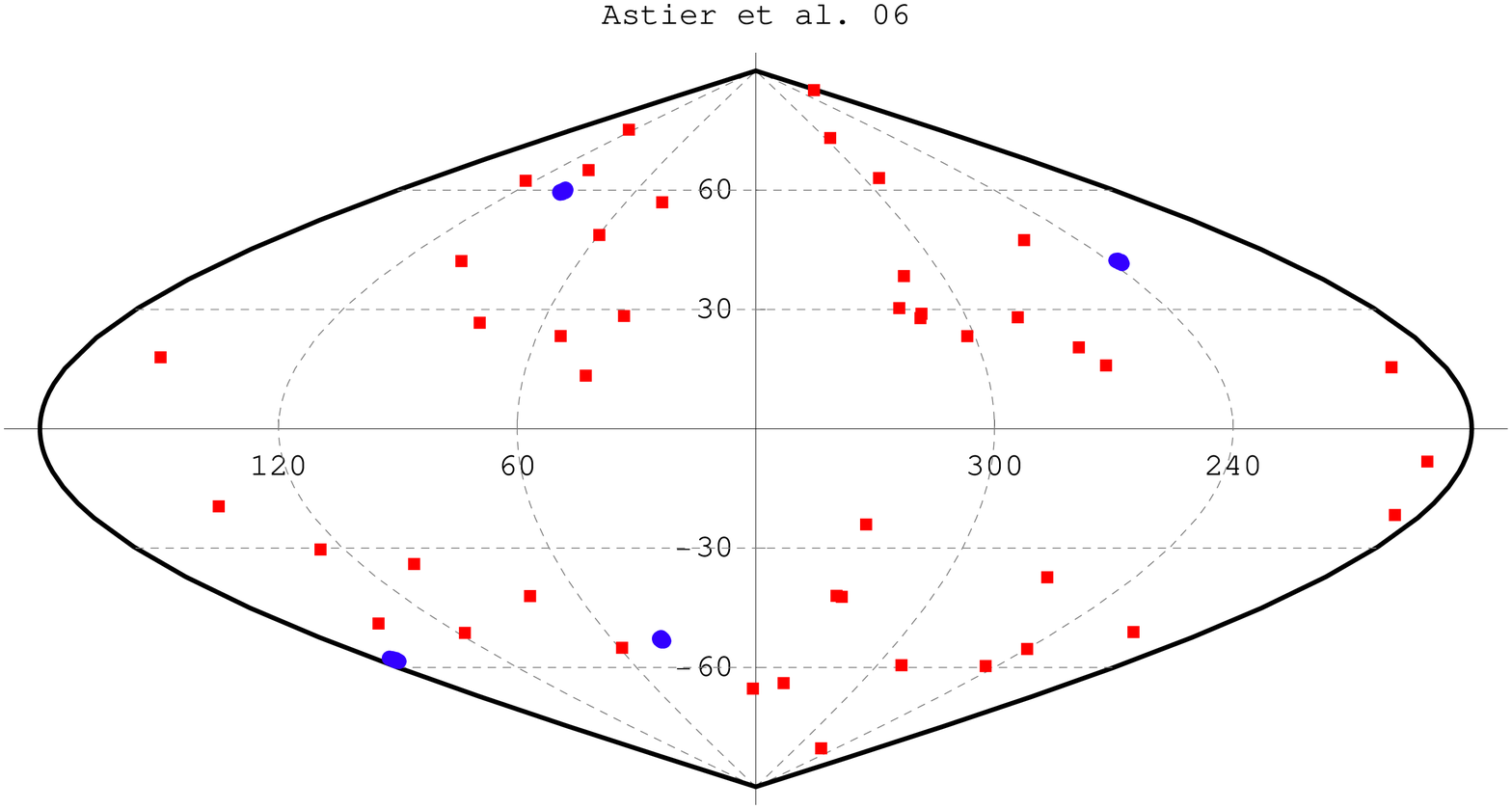}
\includegraphics[width=0.8\linewidth]{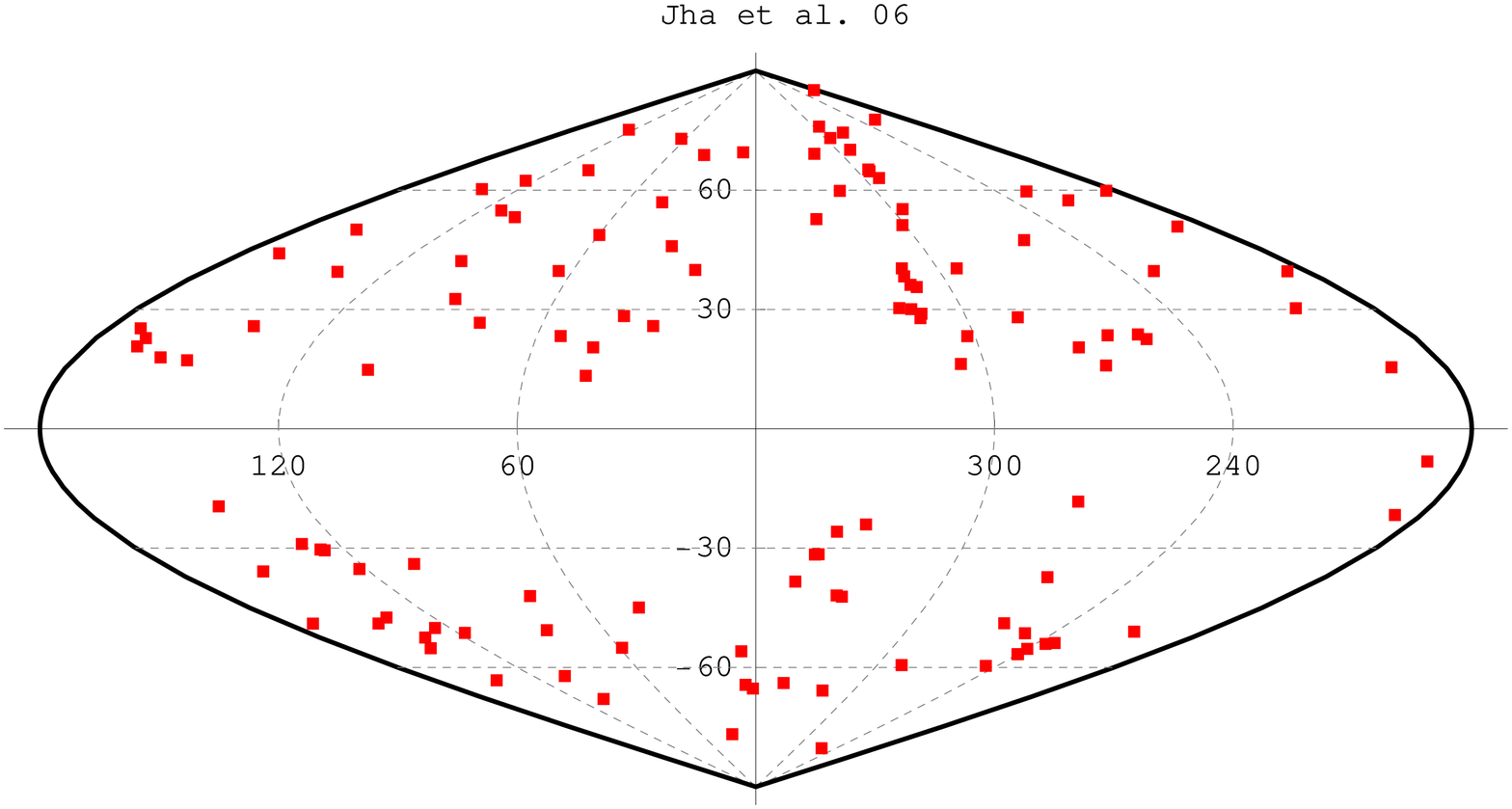}
\caption{\label{fig1}
Distribution of SNe Ia on the sky in galactic coordinates. Red
squares denote SNe with $z<0.2$, whereas blue disks stand for SNe with
$z>0.2$. Only SNe with extinction $A_V < 1$ are shown. For SNe indicated by
open symbols no information on $A_V$ is available.
From top to bottom: 244 (105 with $z < 0.2$, 139 with $z > 0.2$) SNe Ia from
Tonry et al.~(\cite{Tonry:2003zg}) and
Barris et al.~(\cite{Barris:2003dq});
182 SNe Ia (77 with $z < 0.2$ and 105 with $z > 0.2$)
from Riess et al.~(\cite{Riess:2004nr}); 117 SNe Ia
(44 with $z<0.2$ and 73 with $z > 0.2$) from Astier et
al.~(\cite{Astier}), one can easily spot the four SNLS fields of view
(blue);  119 SNe Ia from Jha et al.~(\cite{Jha}) all at small redshift.}
\end{figure}

\begin{figure}
\centering
\includegraphics[width=0.643\linewidth]
{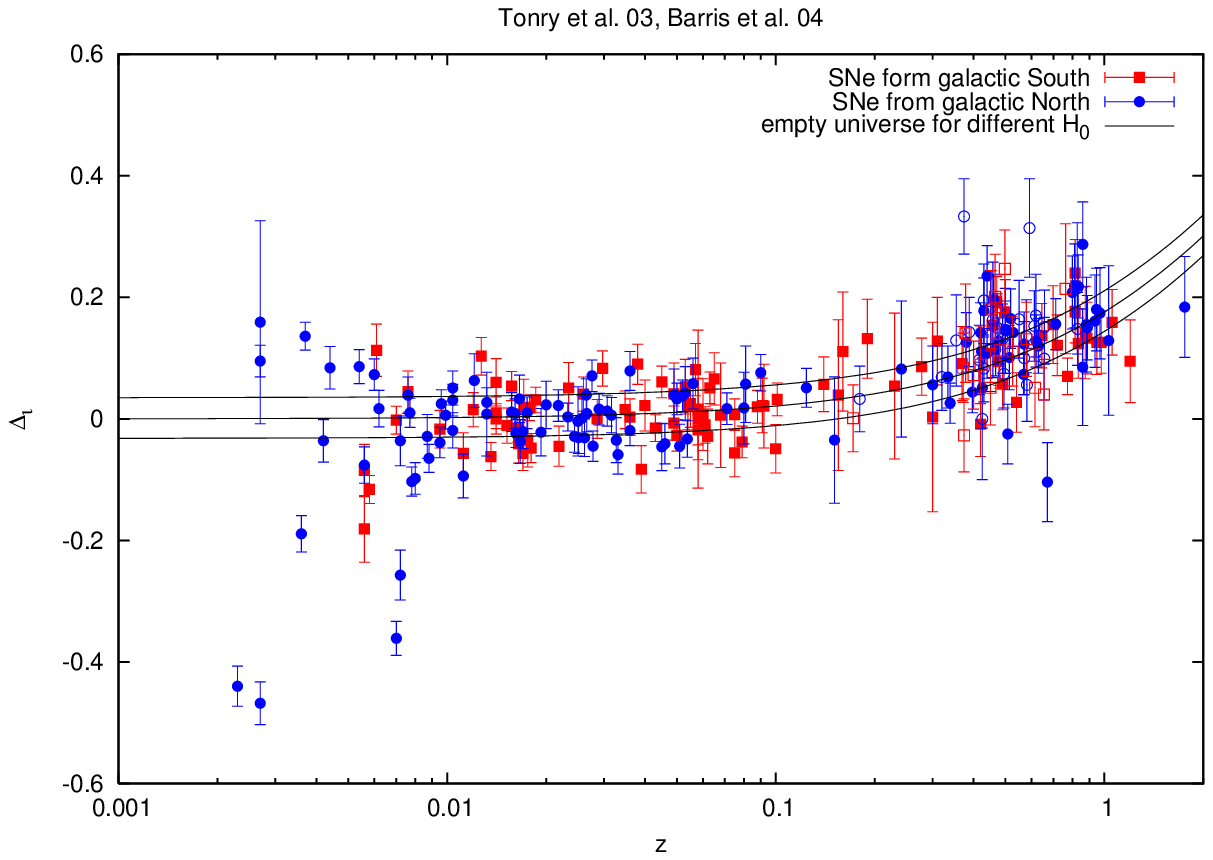}
\includegraphics[width=0.643\linewidth]
{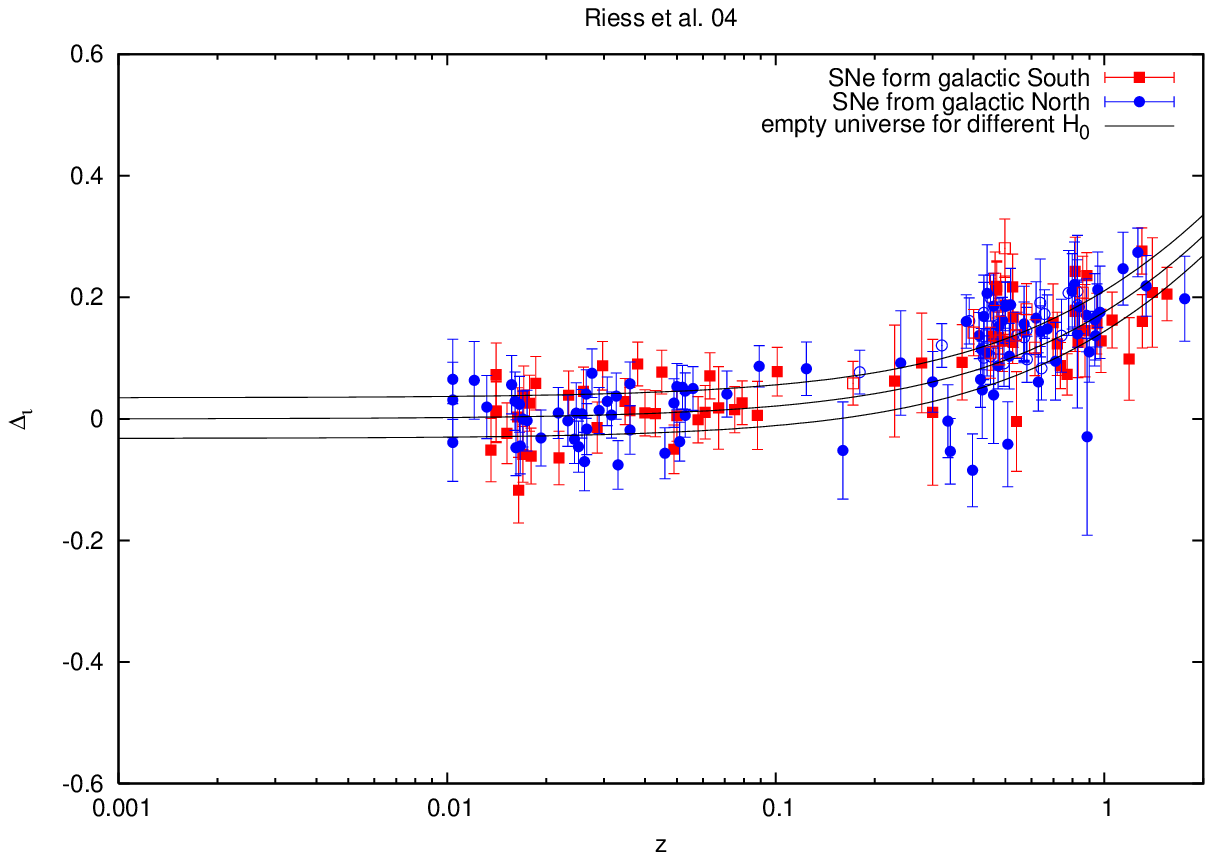}
\includegraphics[width=0.643\linewidth]
{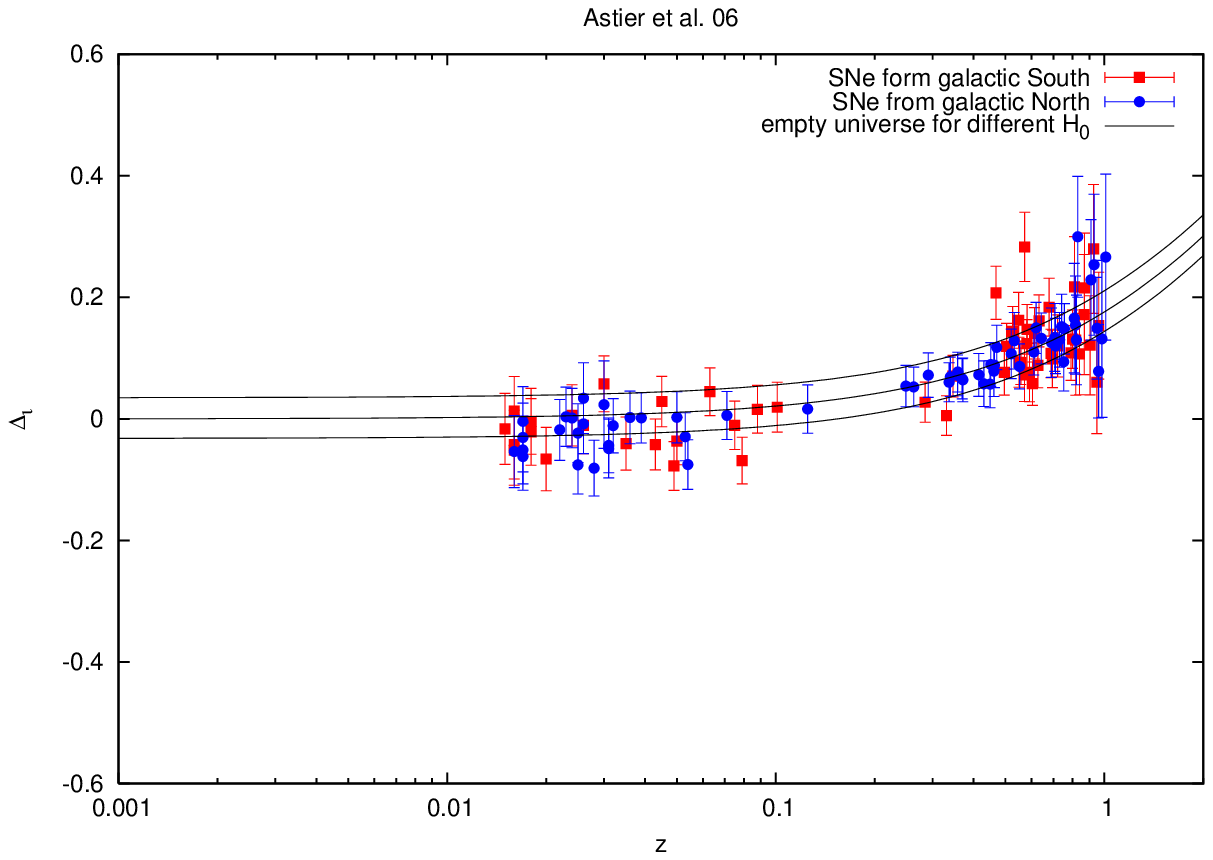}
\includegraphics[width=0.643\linewidth]
{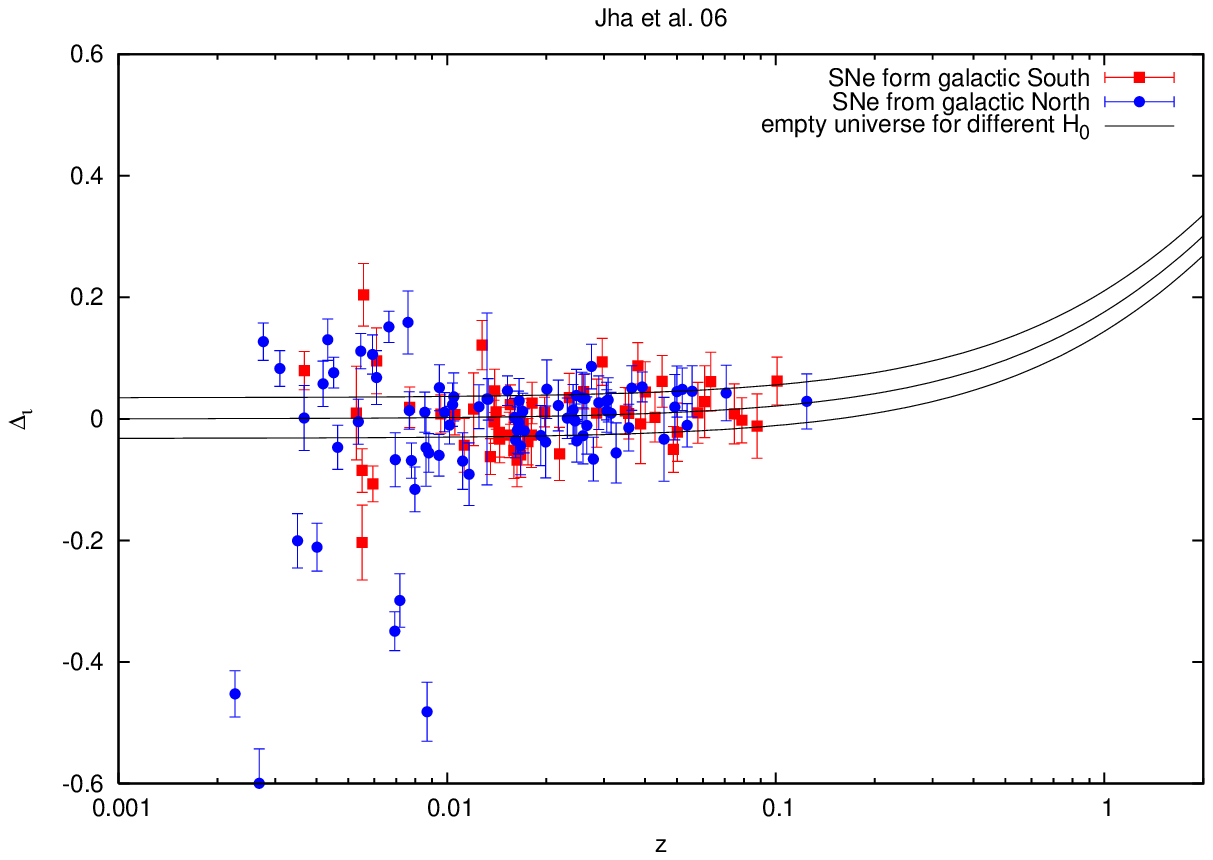} \caption{\label{fig3} Deviation
from the linear Hubble law (see equation \ref{deltai}) for the four
data sets from
figure \ref{fig1}. SNe from the North (South) galactic hemisphere are shown
in blue (red). As a reference we show the expected deviation
for the empty Universe model ($q = j = 0; k = -1$) with three different
values of the SN calibration (black solid lines), corresponding to
$H_0/H_0^* = 0.85, 1, 1.15$ from top to bottom.}
\end{figure}

For the purpose of our model independent analysis we define the deviation
from the linear Hubble law
\begin{equation}
\label{eq6}
\Delta \equiv \log (H_0 d_{\rm L}) - \log (cz_{\rm cmb}),
\end{equation}
where $\log=\log_{10}$. In the following $z_{\rm cmb}$ is denoted by $z$.
Note that often the empty Universe model is plotted as a reference, but
that model assumes already the Einstein equation and we would like to
avoid in our first test any reference to the dynamics of the Universe
and stick to a purely kinematic test combined with the spatial
information provided by the data.

The quantity $\Delta$ is free of the Hubble constant, once
(\ref{HubbleLaw}) or (\ref{Hubble2}) are used. Equivalently,
$\log(H_0 d_{\rm L})$ is related to the ``Hubble-constant-free''
distance modulus
\[
m - {\cal M} = 5 \log(H_0 d_{\rm L}),
\]
with ${\cal M} \equiv M - 5 \log H_0 + 25$
(Perlmutter et al.~\cite{Perlmutter:1998np}). However, to specify
the SN data (in terms of $\log(H_0 d_{\rm L})$ or $m - {\cal M}$)
a certain calibration has to be adopted, which makes use of fiducial
values $H_0^*$ and ${\cal M}^*$.

In order to use SNe for the study of cosmology, one assumes that type
Ia SNe can be made standard candles to a good approximation, i.e.~${\cal M}$
becomes a universal number. As this number depends mainly on the physics
of SNe (the universality of $H_0$ is taken for being granted), it is
treated as a nuissance parameter in the cosmological data analysis.
Consequently, for the estimation of cosmological parameters one marginalises
over ${\cal M}$ (or equivalently $H_0$). However, if we are interested
in effects of large-scale structure formation or a test of the
cosmological principle, we need to study the direction dependent off-set
from the adopted calibration.

Below we follow the convention of Tonry et al.~(\cite{Tonry:2003zg}), where
a Hubble rate of $H_0^* = 65$ km/s/Mpc is assumed;
we thus define (the index $i$ denotes
a particular SN)
\begin{equation}
\label{deltai}
\Delta_i \equiv \log(H_0^* d_{{\rm L}\; i}) - \log(cz_i),
\end{equation}
which is compared to the second order expression
\begin{equation}
\Delta^*(z_i) = - \log(H_0/H_0^*) + \log[1 + (1-q_0)\frac{z_i}{2}],
\end{equation}
or the equivalent in case of the flat $\Lambda$CDM fits. We treat all
data sets (see below) in the same way.

As we have no access to the absolute calibration of the SNe, we cannot
measure $H_0$. However, we can measure the ratio $H_0/H_0^*$, or
${\cal M} - {\cal M}^* = - 5 \log(H_0/H_0^*)$, i.e.~a calibration off-set.
Thus, a $10\%$ effect in $H_0/H_0^*$ corresponds to $0.2$ mag in ${\cal M}$.
The difference ${\cal M} - {\cal M}^*$ seems to be better suited for the
comparison of different data sets, whereas we work with $H_0/H_0^*$,
which seems to us better suited for the study of cosmological
anisotropies.

\section{Data}
\label{data}

We apply our tests on four different data sets that differ in sky and
redshift coverage, shown in figures \ref{fig1} and \ref{fig3},
as well as in systematics. A summary of the four data sets is given in
table \ref{summarytable}.

\begin{table}
\centering
\begin{tabular}{llccc}
 & & full set & $A_V\! <\! 1$ & $A_V\! <\! 1, z\! <\! 0.2$ \\
set & reference(s) & \#SNe $(\bar{z})$ & \#SNe $(\bar{z})$ &
\#SNe $(\bar{z})$\\
\hline A & Tonry et al.~03 & 253 (0.263) & 244 (0.273)
&
139 (0.037) \\
& Barris et al.~04 & & & \\
B & Riess et al.~04 & 186 (0.393) & 182 (0.398) & 77
(0.041) \\
C & Astier et al.~06 & 117 (0.407) & 117 (0.407) & 44 (0.038)
\\
D & Jha et al.~06 & 131 (0.023) & 119 (0.024) & 119 (0.024) \\
\end{tabular}
\caption{\label{summarytable}
Compilation of some characteristics of data sets and their
subsamples used throughout this work.}
\end{table}

Our first data set (A) consists of $253$ SNe
with redshifts $z \in [0.0023,1.755]$ (mean $\bar{z} = 0.27$) from
Tonry et al.~(\cite{Tonry:2003zg}) and Barris et al.~(\cite{Barris:2003dq}).
This data set has a fairly homogeneous sky coverage (except in the zone of
avoidance) and is therefore suitable for our analysis. For each SN Ia we use
the (galactic) coordinates, the distance modulus and the redshift in
the CMB frame. We also use information on light extinction to exclude
supernovae with $A_V > 1.0$ from our analysis. A subset of 42 SNe (originally
from the Supernova Cosmology Project) has no information on $A_V$ specified.
We include the specified errors of the distance modulus and the redshift in
our analysis and treat them as $1\sigma$ errors. The errors on the distance
modulus includes the error from photometry, the scatter between
different analysis methods and an error from the intrinsic dispersion of
SNe Ia.

The second data set (B) is from Riess et al.~(\cite{Riess:2004nr}),
containing $186$ SNe with $z \in [0.0104, 1.755]$ with mean $\bar{z} = 0.40$.
Data set B is not independent from data set A, as it largely contains the same SNe.
However, different corrections have been applied and different selection
criteria have been used. A comparison of data set A and data set B can
thus serve as an estimate of systematic errors in the SNe Ia data
reduction. Data set B contains a `gold' and a `silver' set. We make use
of the `silver' set here in order to have a good sky coverage. It should be
stressed that the number of SNe with $z < 0.2$ in set B is significantly
smaller than in set A.

As a third data set (C) we use the more recent SuperNova Legacy Survey (SNLS)
release (Astier et al.~\cite{Astier}); $117$ SNe, $z = [0.015, 1.01]$ and $\bar{z} = 0.41$.
The SNe at high redshift are independent from sets A and B, but come
from four fields of view only (see figure \ref{fig1}). Thus we should
expect that the pencil beam geometry
of this data set is not ideal for the type of analysis that we have in mind,
although it provides improved control over observational systematics and
has very small errors compared to data set A and B. At low redshifts,
data set C relies largely on the same SNe as data set A and B, but again
with different processing and selection criteria, so it provides again a
cross check for systematics. In contrast to the sets A and B, the intrinsic scatter of
SNe is not included in the error provided by the SNLS team, rather it
has to be incorporated as an additional term in the fits.

Finally, we also use the most recent compilation of  `local'
SNe from Jha et al.~(\cite{Jha}) (data set D);  $131$ SNe, $z \in [0.00226, 0.1242]$ and
$\bar{z} = 0.024$. It largely contains the same SNe as data set A, the main
difference being a largely improved procedure to obtain the distance moduli
and their errors. This set has a fairly good sky coverage, but is limited to small redshifts.
Like for data set C the intrinsic scatter is not included in the provided error.

Two more recent data sets are not used because they contain SNe from
very small regions of the sky only, i.e.~from the two HST GOODS
fields (Riess et al.~\cite{Riess06}) and 20 fields of view within
three hours in right ascension (ESSENCE collaboration: Wood-Vasey et
al.~\cite{WV07}). These surveys suffer from the same limitations as
the SNLS data, they are a set of pencil beams.

For all data sets we can see from figure \ref{fig1} that the sky
coverage is much better for the low redshift SN ($z < 0.2$).
Figure 2 provides an impression of the redshift distribution of the
samples for both galactic hemispheres. The current lack of SN observations at
intermediate redshifts ($z = 0.1$ to $0.3$) is easily spotted. At small
$z < 0.01$ the spreading of the distribution due to peculiar velocities
of the SNe is seen. The deviation from the linear Hubble law becomes
visible at $z > 0.1$.

The four data sets used here differ from each other
in the method applied to estimate the values of the fitted
parameters. In order to fit the SN observations of set A, Tonry et al.~(\cite{Tonry:2003zg})
and Barris et al.~(\cite{Barris:2003dq}) used four fit methods: the MLCS method
described in Riess et al.~(\cite{Riess:1998cb}), the $\Delta m_{15}$ method
(Philips et al.~\cite{phillips}) and its improved version dm15 (Germany
et al.~\cite{germany}). Additionally, they used the BATM method by
Tonry et al.~(\cite{Tonry:2003zg}). For data set B, the BATM method
was used to check the results with the improved version MLCS2k2 described
in Jha~(\cite{jha2002}) and Jha et al.~(\cite{Jha}). This method is
also used by Jha et al.~(\cite{Jha}) to fit the SNe of data set D. Astier et al.~(\cite{Astier})
used the SALT method (Guy et al.~\cite{guy}) to fit the data of set C.

The reader should be aware that Tonry et al.~(\cite{Tonry:2003zg}),
Barris et
al.~(\cite{Barris:2003dq}) and Riess et al.~(\cite{Riess:2004nr}) provide
$z_{\rm cmb}$ for each SN,
whereas Astier et al.~(\cite{Astier}) and Jha et al.~(\cite{Jha})
give $z_\odot$.

\begin{figure}
\centerline{
\includegraphics[width=0.8\linewidth]{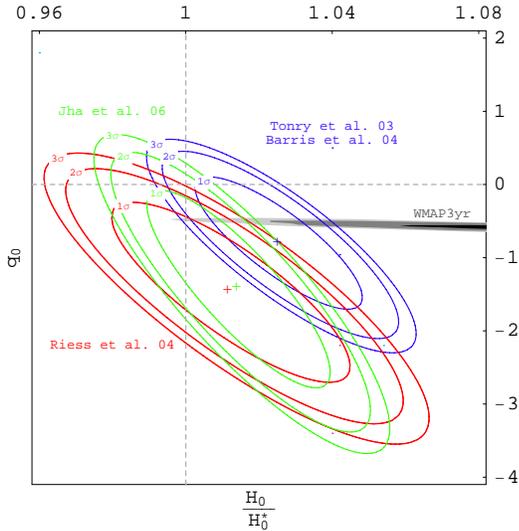}}
\caption{\label{IFSHoqoTBRA} Confidence contours for a model-independent
full-sky fit to the Hubble law at second order for three SNe Ia data
sets. SNe up to redshift $z = 0.2$ are included in the fits. We determine
the calibration off-set $H_0/H_0^*$ and the deceleration parameter $q_0$.
As a comparison we show the corresponding contours of the WMAP measurement
(Spergel et al.~\cite{Spergel:2006hy}). In contrast to the spread
of the SN calibration off-set, its best-fit value has no physical
relevance in the case of full sky fits. Thus the WMAP data may be moved
horizontally in this figure.}
\end{figure}

\section{Comparing hemispheres}
\label{method}

As motivated in the introduction, we wish to test the isotropy of the
Hubble diagram. As the number of observed SNe is still small, it does
not make sense to look into small-scale variations, rather we should
look at the largest possible scales. The largest possible anisotropy
scale is $180^\circ$. A very simple test is splitting the sky into
hemispheres and to compare the corresponding Hubble diagrams.
We can search for directions of anisotropies by rotating the respective
poles over the sky. A similar study has been used by Eriksen et al.~(\cite{Eriksen})
to analyse the cosmic microwave sky, which led to one of the several anomalies
discovered in the WMAP data (see Copi et al.~\cite{Copi} for a recent update and
summary).

\begin{figure*}
\centerline{
\includegraphics[width=0.33\linewidth]{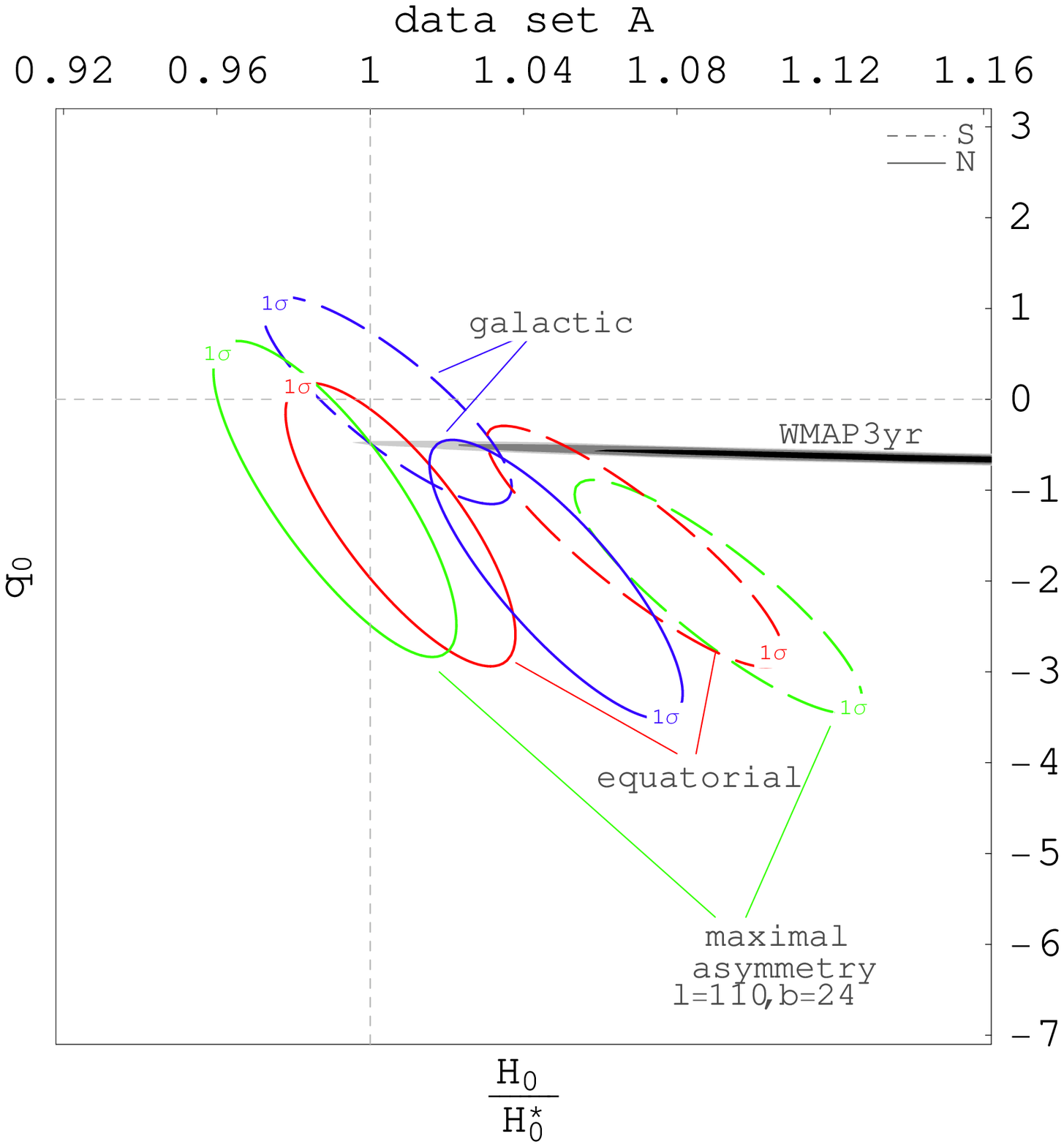}
\includegraphics[width=0.33\linewidth]{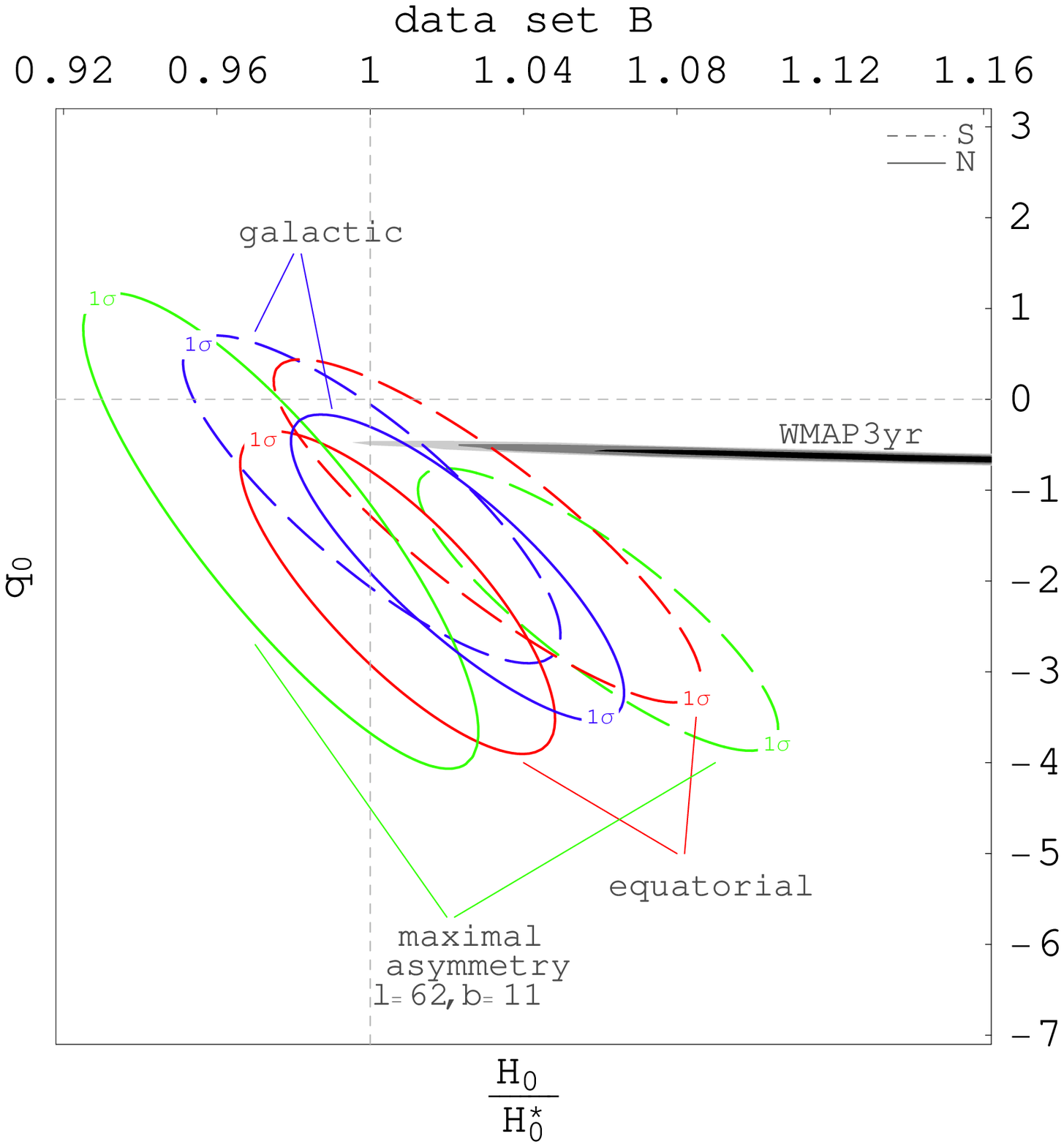}
\includegraphics[width=0.33\linewidth]{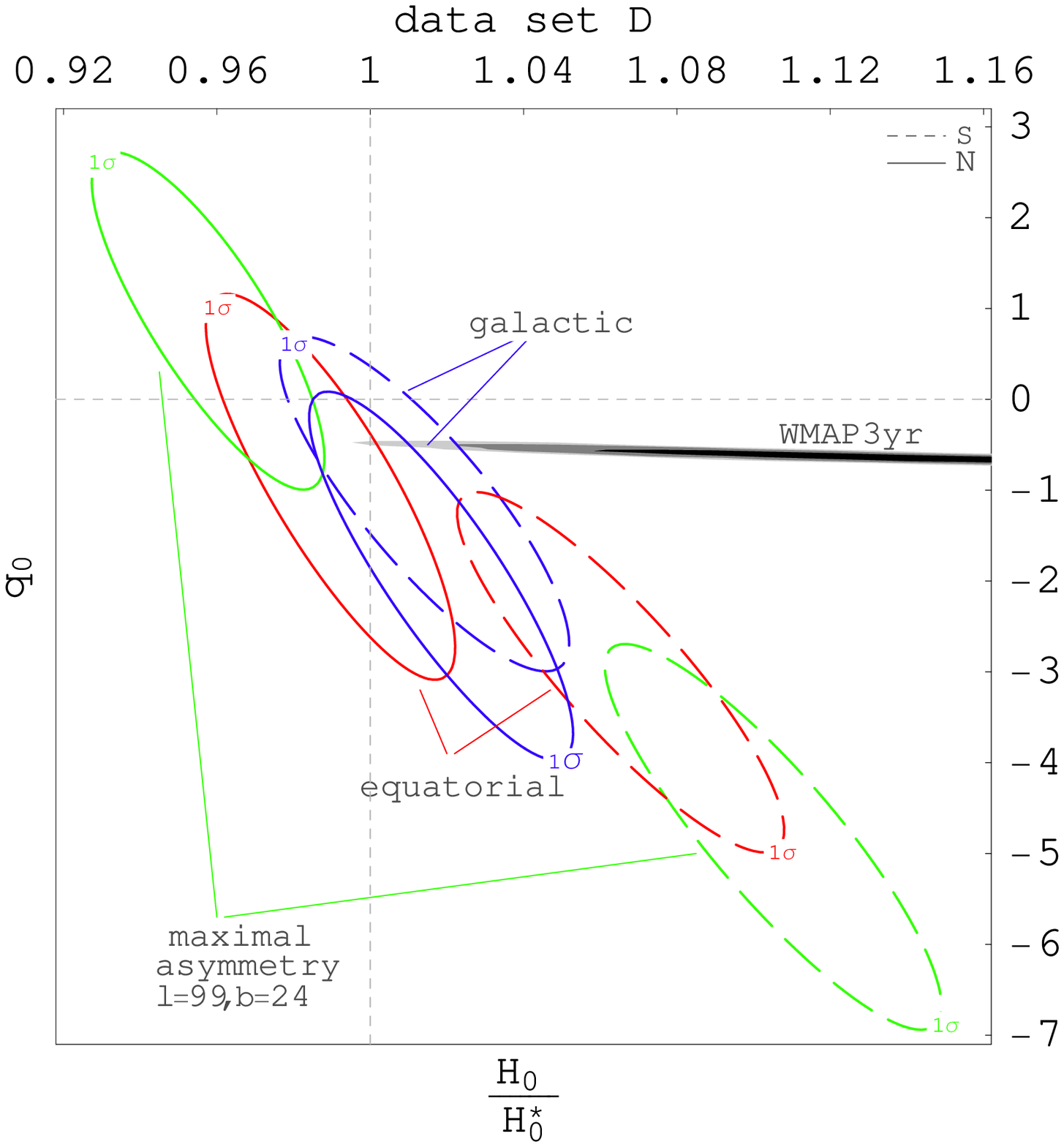}}
\caption{\label{IHoqoTBRgaleqma} North (full lines) and South (dashed lines)
confidence contours and best-fit values for galactic, equatorial and maximum asymmetry
hemispheres for the model-independent test.
These fits should be compared to the full-sky fits of figure
\ref{IFSHoqoTBRA}. We do not show results for data set C, as the number
of SNe at $z<0.2$ of that data set turned out to be insufficient for this type of test.}
\end{figure*}

\subsection{Model-independent test of isotropy}

If we restrict the analysis to SNe at $z<0.2$, it is reasonable to
fit the Hubble law (\ref{HubbleLaw}) and its quadratic correction. This
provides a model independent test of the cosmological principle and at
the same time a check for systematic errors in the SN Ia search,
observation and analysis.

In order to find the best-fitting values of
the calibration $H_0/H_0^*$ and the deceleration $q_0$,
we follow the standard approach and minimise
\begin{equation}
\label{chi2}
\chi^2(H_0/H_0^*,q_0) = \sum_i
\frac{\left[\Delta_i - \Delta^* (z_i; H_0/H_0^*, q_0) \right]^2}
{\sigma_{\Delta_i}^2 + \sigma_{\rm int}^2 +
(\sigma_v^2 + \sigma_{z_i}^2)/z^2_i},
\end{equation}
where the sum goes over a subset of one of the three data sets. For the
data sets A and B the intrinsic dispersion $\sigma_{\rm int}$ is included in
$\sigma_{\Delta_i}$, for data set C and D $\sigma_{\rm int} > 0$. $\sigma_v$ and
$\sigma_z$ denote the dispersion of the peculiar velocities and error of
the redshift measurement. The latter two are most relevant for nearby SNe.

\subsubsection{Full sky}

As a reference we first determine the best fit values for $H_0/H_0^*$ and
$q_0$ for the four SN Ia data sets. In doing so we aim at finding
the best suited values for redshift and light extiction cuts, and we
look a the best suited value of the peculiar velocity dispersion. We
adopt a value of $\sigma_v = 345$ km/s for all four data sets, as it
provides a reasonable $\chi^2$ per degree of freedom of order
one for all data sets under consideration.
The $z < 0.2$ samples from sets
A to D have mean redshifts of $0.038, 0.041, 0.038$ and $0.024$, respectively.

The results of the full-sky fits are presented in figure~\ref{IFSHoqoTBRA} (except for set C)
and in table \ref{TFSHoqo}. The fits shown in the figure correspond to the first
fit in the table for the respective data set. We find that all four data sets provide fits that
are consistent with each other. The largest difference is observed with respect to the
SN calibration ($H_0/H_0^*$) in data set C. This happens because the SN calibration
of that set  has been fit to the WMAP measurement of $H_0$, whereas the other three data
sets did not assume the WMAP value.
At small redshifts, only data set B gives rise to
a marginal evidence for acceleration. However, all data sets are consistent
with acceleration and with the fit to WMAP data.

In order to test the robustness of the fits we vary the value of the
parameters $A_V$, $\sigma_v$, $\sigma_{int}$ and the range in redshift
(see table~\ref{TFSHoqo}). The most significant effect comes from a
restriction of the redshift range to $0.02 < z < 0.2$ or even $0.01<z<0.1$,
which
leads to a noticeable change in the best-fit value for $q_0$ and a substantial
increase in its error. Shifting the maximal redshift to $z=0.3$ changes the
result only slightly. From this point of view it is reasonable to include all SNe with $A_V < 1$
up to $z=0.2$ in order to fit the Hubble law up to the second order in redshift.

Although the full-sky fits indicate a robust data set C, our studies
have shown that it is not well suited to be used within this work
due to its small number of SNe and its bad sky coverage. We
therefore focus on the results of data sets A, B and D in the
following chapters.

\subsubsection{Galactic and equatorial hemispheres}

We start to study the anisotropy of the Hubble diagram by comparing
those hemispheres which have a natural origin such as the galactic
hemispheres and the equatorial hemispheres. The galaxy defines the
zone of avoidance and is approximately North-South symmetric. Thus
the galactic hemispheres cover approximately the same observable regions
of the sky and should be well suited for a cosmological test. On the other hand,
most observations (especially at small redshifts) are done from the ground and thus the
equatorial system is distinguished. Any correlation with this system would hint to
a systematic effect in the search, observation or data analysis of SNe Ia.

Figure \ref{IHoqoTBRgaleqma} shows the 1$\sigma$ contours for the
galactic and equatorial hemispheres. As expected, all three data sets show
a consistent fit for North and South galactic hemispheres. However, the equatorial
system shows some unexpected deviations. For data set A we find a significant
deviation of the calibration $H_0/H_0^*$ for the equatorial hemispheres.
The northern hemisphere favours a lower value for $H_0/H_0^*$.
The same trend is observed in data set B, but from the analysis of data set B alone
we would not pay attention. For data set D the deviation is again clearly seen and is
not restricted to the calibration. The corresponding best fit values and their errors are
displayed in table~\ref{TgaleqHoqo}.

In order to quantify the evidence for an equatorial North-South asymmetry
of the SN calibration, we use Monte Carlo (MC) simulations to check for artefacts of
sky coverage.  We test against 500 random realisations of the same data sets
in which we mix the coordinates l and b for all SN Ia at $z<0.2$.
For each simulated hemisphere pair we calculate the deviation in $\chi^2$ as
\[
\Delta \chi^2 \equiv
\mbox{min} \left(|\chi^2_N(S) - \chi^2_N(N)|,  |\chi^2_S(N) - \chi^2_S(S)|\right),
\]
where $\chi^2_N(S)$ denotes the value of expression (\ref{chi2}) for the best-fit parameters
of the North hemisphere applied on the data of the South hemisphere, etc.
We also calculate the deviation in $H_0/H_0^*$ and $q_0$ of the best-fit values
from the opposite hemispheres. The results of these hemispherical fits are given
in table \ref{TgaleqHoqo}.

While no deviation from isotropy is found for data set B,
we find data set A shows a statistically significant anisotropy in the equatorial system.
Only $0.8\%$ of our MCs give rise to a larger difference in $\Delta \chi^2$ and
only $4.6\%$ of the MCs show a larger difference in the calibration
$H_0/H_0^*$. The latter number for data set D is $5.0\%$.  Thus, it seems the equatorial
hemispheres do not agree with the expectation from the cosmological principle at the
$95\%$~C.L. with respect to the SN calibration (evidence from sets A and D).

\subsubsection{Hemispheres of maximal asymmetry}

We found evidence for a significant asymmetry in the equatorial coordinates in data sets A
and D (and consistent with data set B).  We may ask if the asymmetry is maximised by the
equatorial system or if more asymmetric directions exist on the sky. In order to test this
anisotropy for its size we are going to identify those hemispheres
which have the largest deviation in $\chi^2$.

We search for the maximal asymmetric hemispheres by rotating the poles
across the sky. We calculate the deviation in $\chi^2$ for each
hemisphere with poles at $l\in[0^{\circ},180^{\circ}]$ and $b\in[-90^{\circ},90^{\circ}]$
in $1^{\circ}$ steps.

\begin{figure}
\includegraphics[width=0.9\linewidth]
{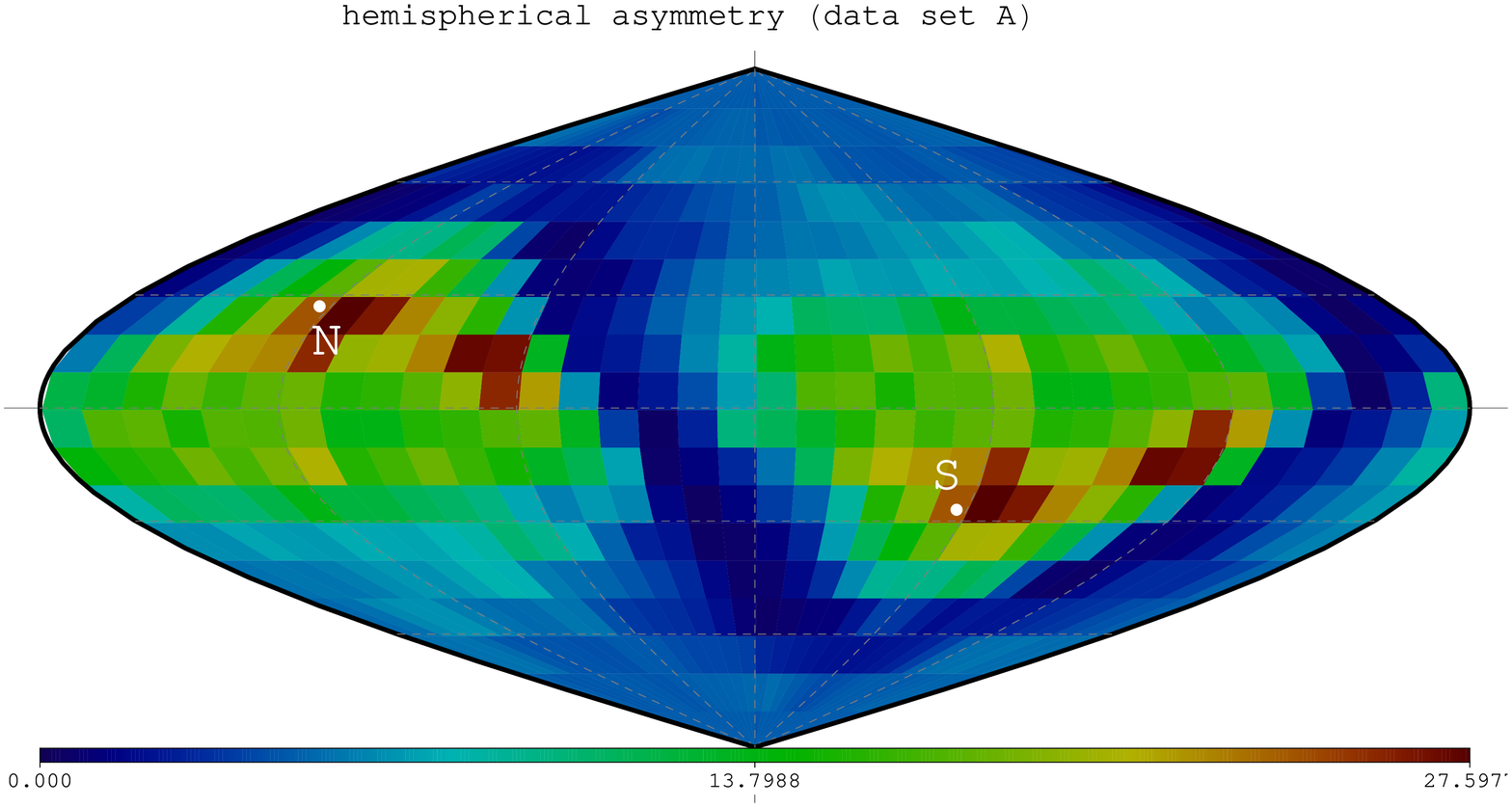}
\includegraphics[width=0.9\linewidth]
{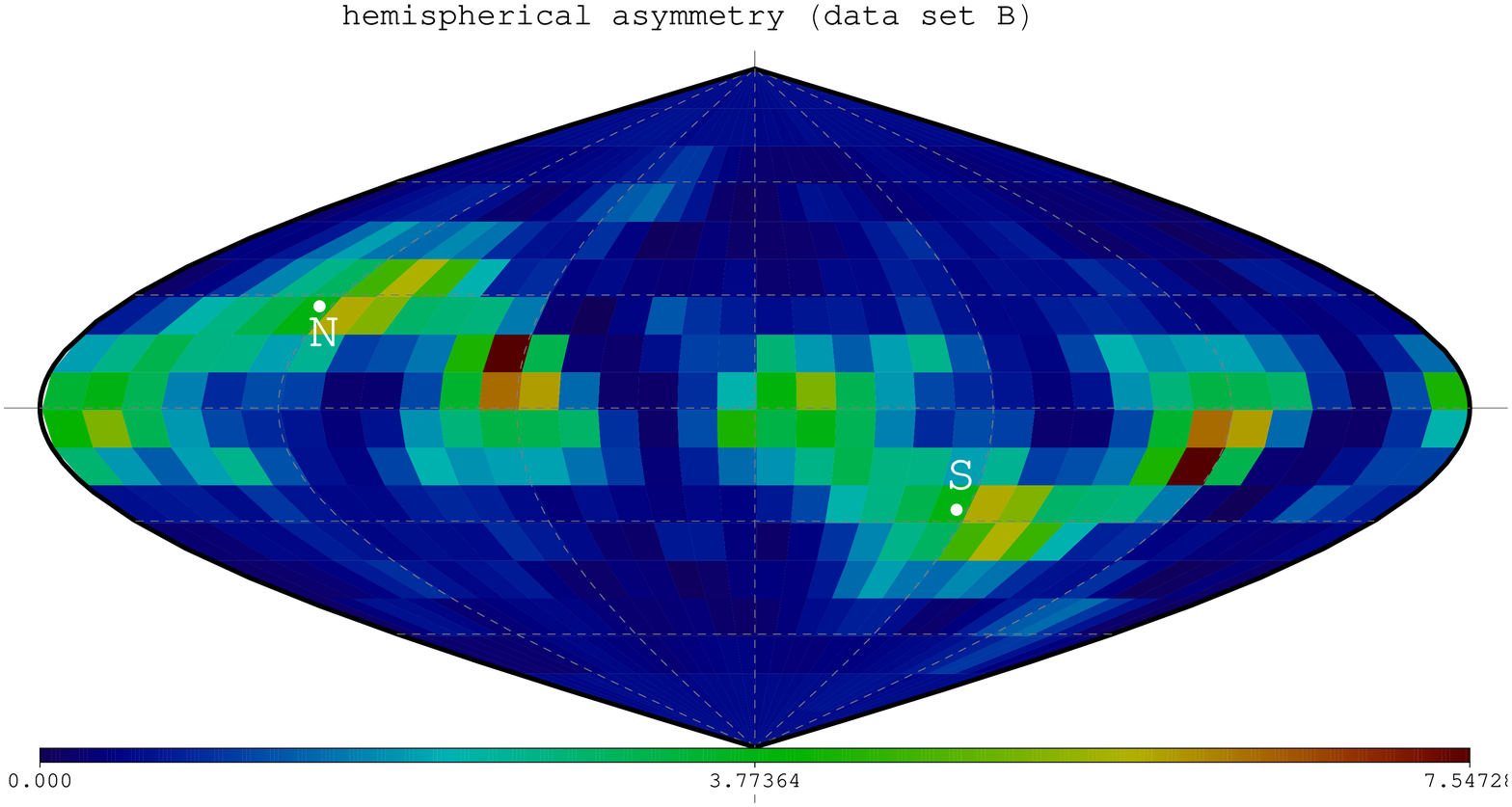}
\includegraphics[width=0.9\linewidth]
{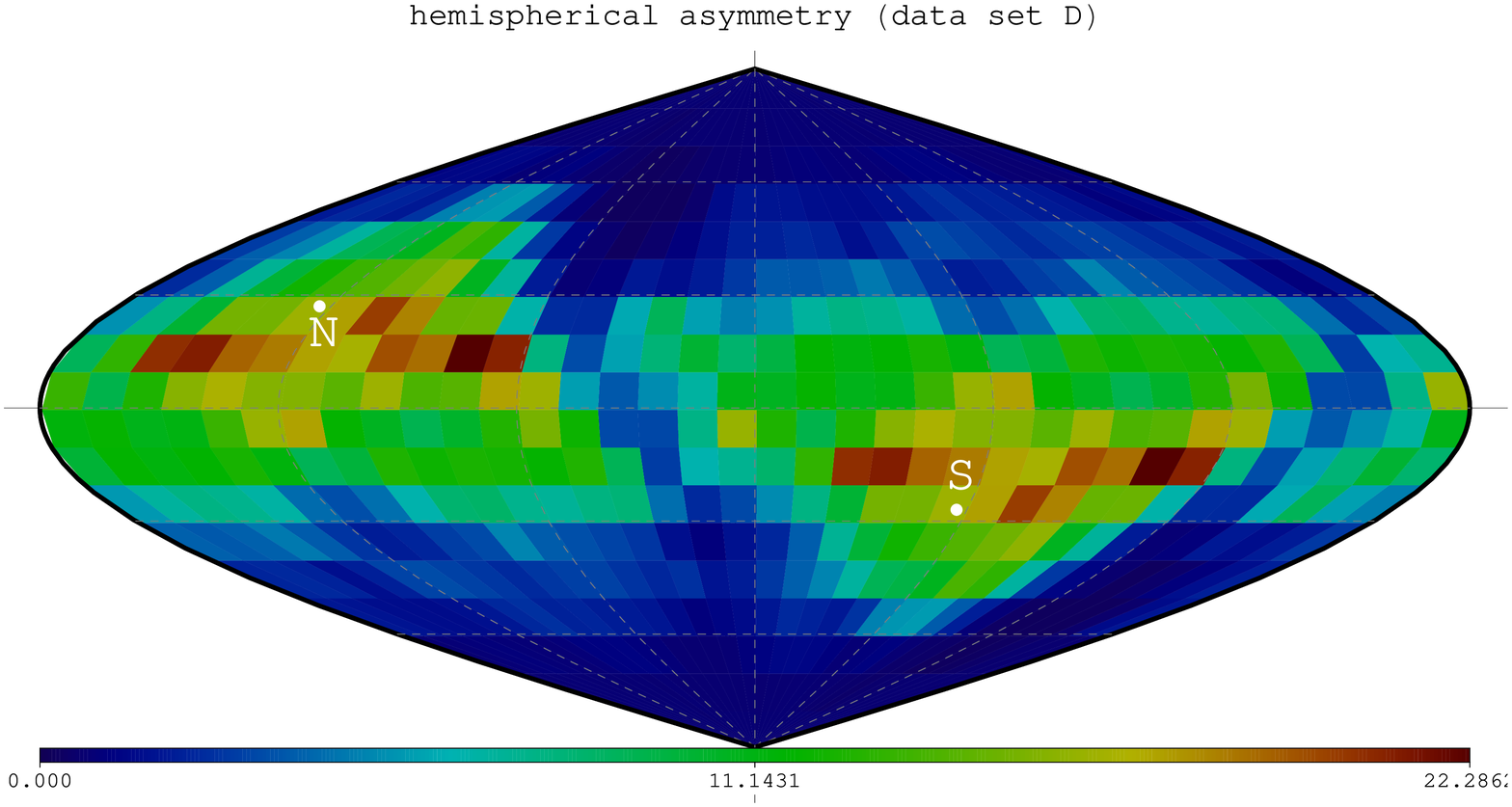} \caption{\label{fig5}
Hemispherical asymmetry in $\Delta\chi^2$ for three SN Ia data sets
at small redshift ($z < 0.2$) for the model independent fit. When
the pole lies in a region with a bluish colour the asymmetry is
small, the red spots denote the directions of large asymmetry. We
plot directions in galactic coordinates. The white points indicate
the zenit (equatorial North) and its antipode (equatorial South).}
\end{figure}

The results are given in table \ref{TgaleqHoqo} and the deviations in $\Delta \chi^2$ are
illustrated in Fig. \ref{fig5}. Each coloured field in the figure is $10^{\circ} \times 10^{\circ}$
in size and its colour represents the mean deviation of the $100$ corresponding pole
positions. It is striking that all three data sets give rise to the same pattern on the sky (especially
sets A and D), despite the fact that the methods of SN data reduction and selection criteria differ.
By construction the pattern is symmetric and thus all information is contained in one
hemisphere. In our discussion below we refer to the North galactic hemisphere.

The asymmetry is strongest in data set A, followed by D and B. As data set A contains more
SNe than set D, which contains more than set B, it seems that this trend is in accordance with the statistical power of the data sets. One can see, that two asymmetric directions are common
in all three data sets. The first one is close to the equatorial poles at $(l,b) = (123^\circ,27^\circ)$,
the other one is close to $(l,b) = (70^\circ, 15^\circ)$. The proximity to the equatorial system is most
pronounced in set A. Sets B and D maximise the asymmetry toward a direction  $\sim
(70^\circ, 15^\circ)$.

Let us also note that the observed pattern is unexpected. For randomly distributed SNe
with a gaussian scatter in magnitude, we would expect that the resulting pattern would show
less structure.

We performed 500 MC simulations in order to check if the observed amount of asymmetry
is to be expected for a data set with this kind of sky coverage. As above we mix the
coordinates of the SNe of the data set under consideration and search
for the maximal asymmetry in each of the simulated data sets. In order to keep
the computational effort to a minimum, we now use a step size of $5^\circ$. The
asymmetry of the MC data sets is compared to the one from the original set. The results
are given in table \ref{tab4}. The larger step size in the search for the maximal asymmetric
direction explains why the directions of maximal asymmetry differ from the ones in table
\ref{TgaleqHoqo}. For data sets A and D the asymmetry is larger than in $99.8\%$ of our MCs.
The deviation in the SN calibration seems to be significant at $> 96\%$~C.L. Data set D shows
also a significant asymmetry in the extracted values of the deceleration parameter. Only
$1.8\%$ of the MCs show a larger difference.
In contrast,  the asymmetry in data set B appears not to be statistically significant. However,
we should keep in mind that this data set shows a very similar asymmetry pattern but contains
less SNe than sets A and D.

\subsubsection{Summary of model-independent test}

For data set A we have identified a statistically significant asymmetry in the equatorial
system. It turns out to be due to an off-set in the calibration $H_0/H_0^*$ among
the two hemispheres. The direction of maximal asymmetry is very close to that
direction in data set A. Data set B does not show the same amount of asymmetry, but is
qualitatively consistent with data set A. However, it contains less SNe and it is thus expected
that the asymmetry should be less obvious. The larger data set D shows again a statistically
significant asymmetry in the SN calibration.

We cannot offer an explanation for the asymmetry, but we think that the proximity of the
direction of maximal asymmetry to the equatorial poles suggests an systematic error in
one of the steps (search, observation, data analysis) to the calibration of SNe.

A second direction of asymmetry has been identified in all three data sets. Its detection is statistically
less significant.

\begin{figure}
\centerline{
\includegraphics[width=0.9\linewidth]
{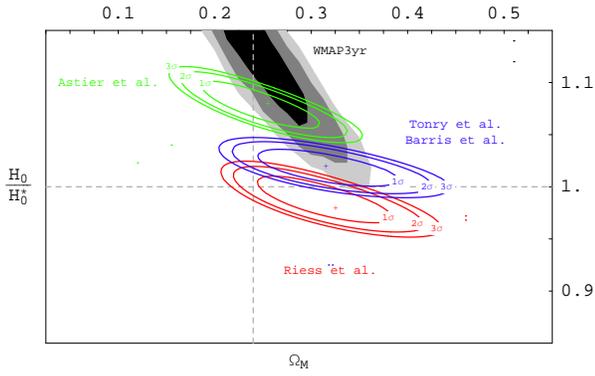}} \caption{\label{OmH0TBRJ}
Confidence contours for a full-sky fit to the flat $\Lambda$CDM
model for three SNe Ia data sets. All SNe with $A_V < 1$ are
included in the fits. We determine the calibration off-set
$H_0/H_0^*$ and the dimensionless mass density $\Omega_{\rm M}$. For
comparison we show the corresponding contours of the WMAP
measurement (Spergel et
al.~\cite{Spergel:2006hy}). For the latter $H_0$ indeed is the
extracted Hubble parameter. Here the best-fit value of
the calibration off-set is of no relevance, the vertical extension
of the contours however reflects the physical scatter of the data.}
\end{figure}

\subsection{Isotropy in the flat $\Lambda$CDM model}

Let us now include SNe Ia at any redshift in our test. This allows us to the test the
concordance model of cosmology. We restrict our analysis to the flat $\Lambda$CDM
model. Thus, the Hubble diagram can now be used to fit $\Omega_{\rm M}$ and $H_0/H_0^*$,
the calibration off-set of SNe Ia. Here we do not make use of data set D, as it is limited to small
redshifts.

\subsubsection{Full sky}

To start with, we show the results for the full-sky fit in table \ref{tabLCDMfull}.
Again we see that the fits are quite robust, except for the restriction to $z<0.2$.
In that case, a matterless Universe ($\Omega_{\rm M} = 0$) cannot be ruled out.
The confidence contours for the $\Lambda$CDM fits to SNe Ia at arbitrary redshift
are shown in figure \ref{OmH0TBRJ}. Only for data set C the best fit value is close
to the one of the WMAP measurement. This is due to the fact that the SN calibration in
set C was chosen to agree with WMAP measurements, whereas for SNe sets A and B
no information from the CMB was used. In the following we restrict our presentation
to data sets A and B, as the pencil beam geometry of data set C turned out not to
be suitable for our tests.

\begin{figure}
\includegraphics[width=0.9\linewidth]
{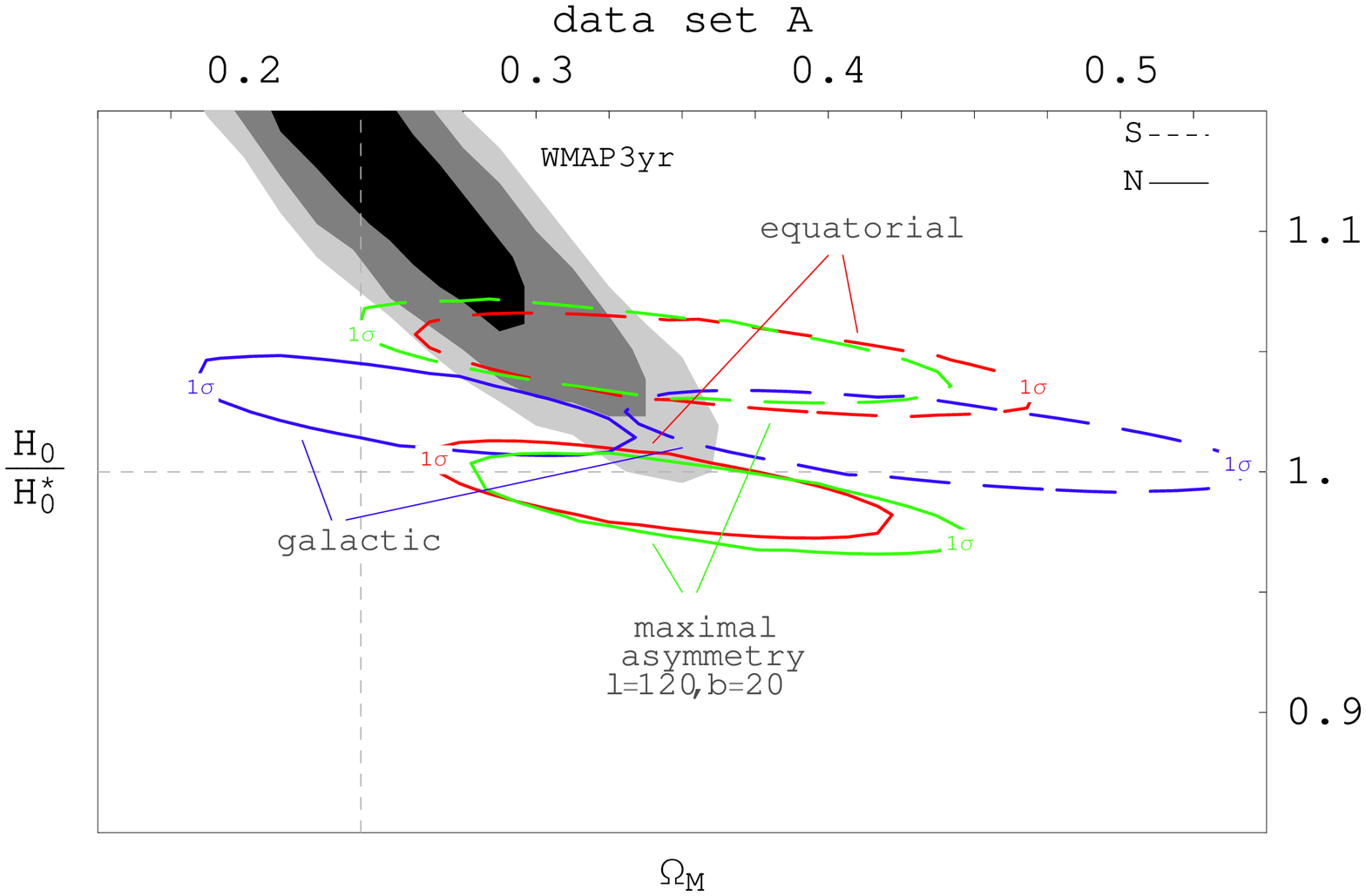}
\includegraphics[width=0.9\linewidth]
{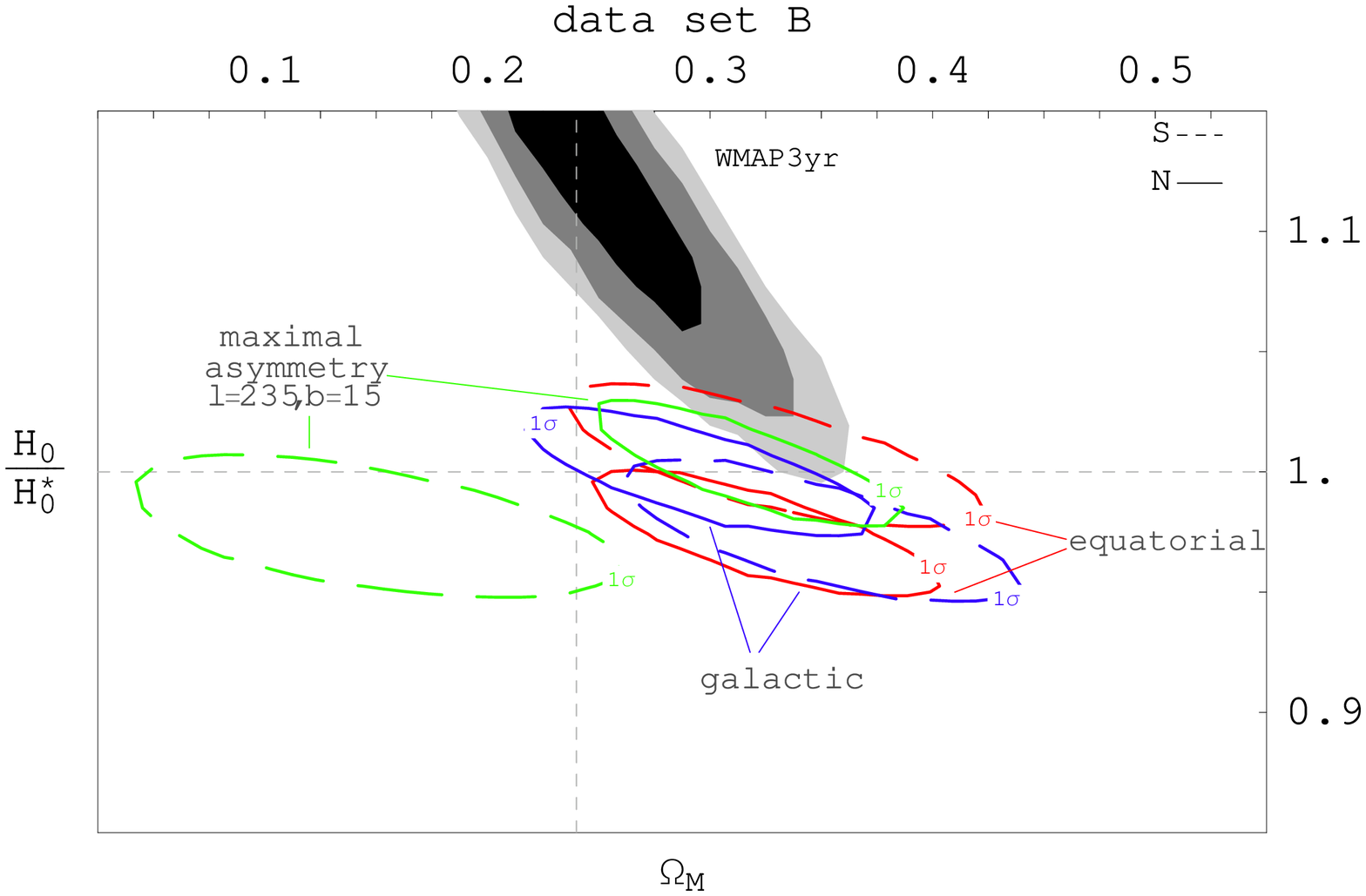} \caption{\label{fig7} North
(full lines) and South (dashed lines) confidence contours and
best-fit values for galactic, equatorial and maximum asymmetry
hemispheres for the $\Lambda$CDM fit. These fits should be compared
to the full-sky fits of figure \ref{OmH0TBRJ}. We do not show
results for data set C, as the pencil beam geometry of that data set
is not suitable for our test.}
\end{figure}

\subsubsection{Galactic and equatorial hemispheres}

The fits to North and South galactic and equatorial hemispheres are presented in
table \ref{LCDMgaleqHoqo}. The asymmetry in equatorial coordinates found
at low $z$ is confirmed by the full data sets A and B. Again we run 500 MCs to
test the statistical significance and find that the calibration off-set in the equatorial
coordinates is now significant at $> 95\%$~C.L. in {\emph both} data sets. In data set A
it is at $> 99\%$~C.L. The corresponding confidence contours are shown in figure \ref{fig7}.

Additionally, we find now that data set A shows also asymmetry in galactic coordinates, which
is an asymmetry in the extracted matter density at $95\%$~C.L.; see table \ref{LCDMgaleqHoqo}
for the strongly deviant values of $\Omega_{\rm M}$ in the North and South hemispheres,
$|\Delta \Omega_{\rm M}| = 0.2$.  This asymmetry is not at an significant level in data set B, but as
for the model-independent test, data set B is fully consistent with the conclusions from data set A.

\subsubsection{Hemispheres of maximal asymmetry}

\begin{figure}
\includegraphics[width=0.9\linewidth]
{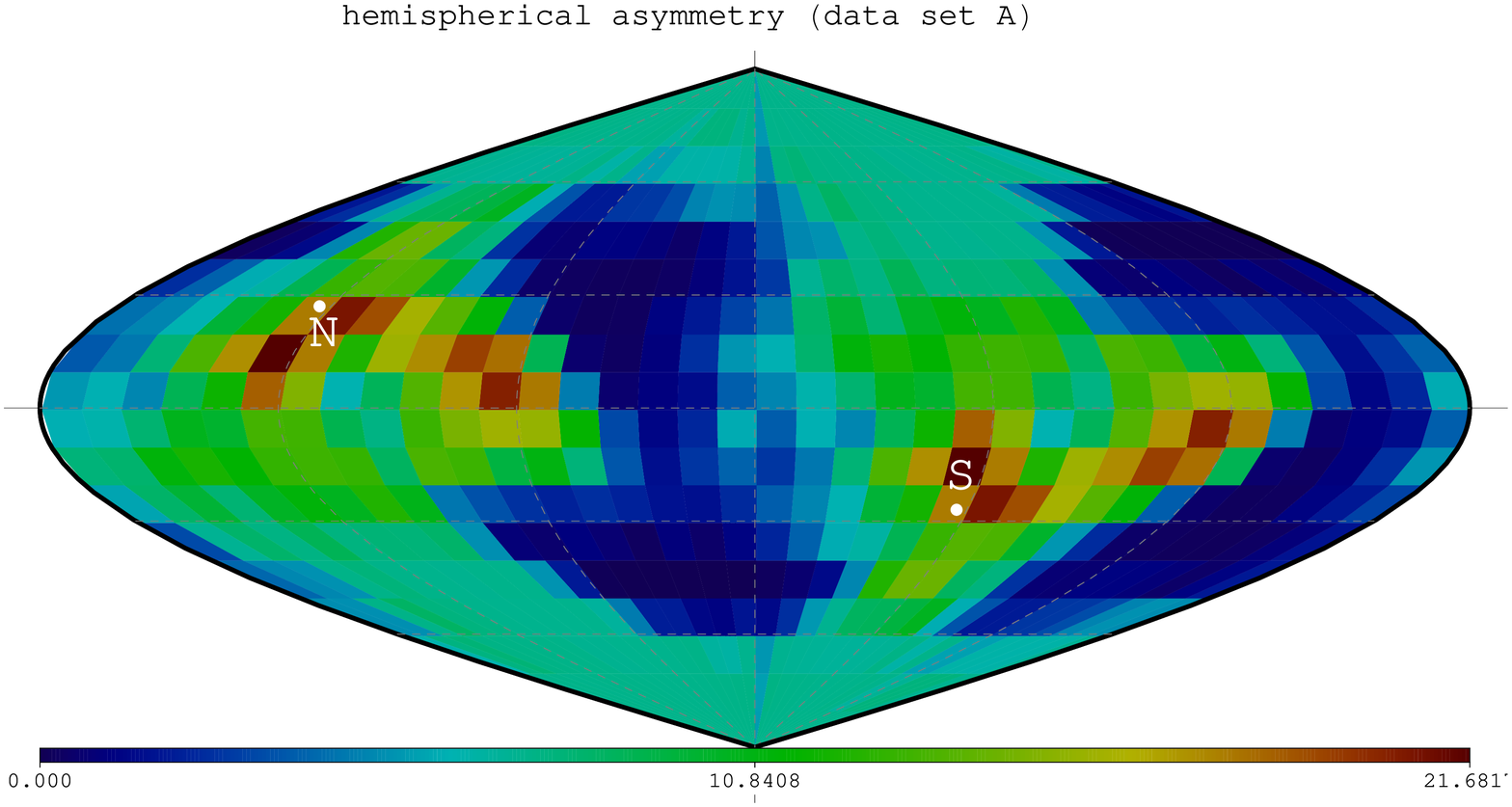}
\includegraphics[width=0.9\linewidth]
{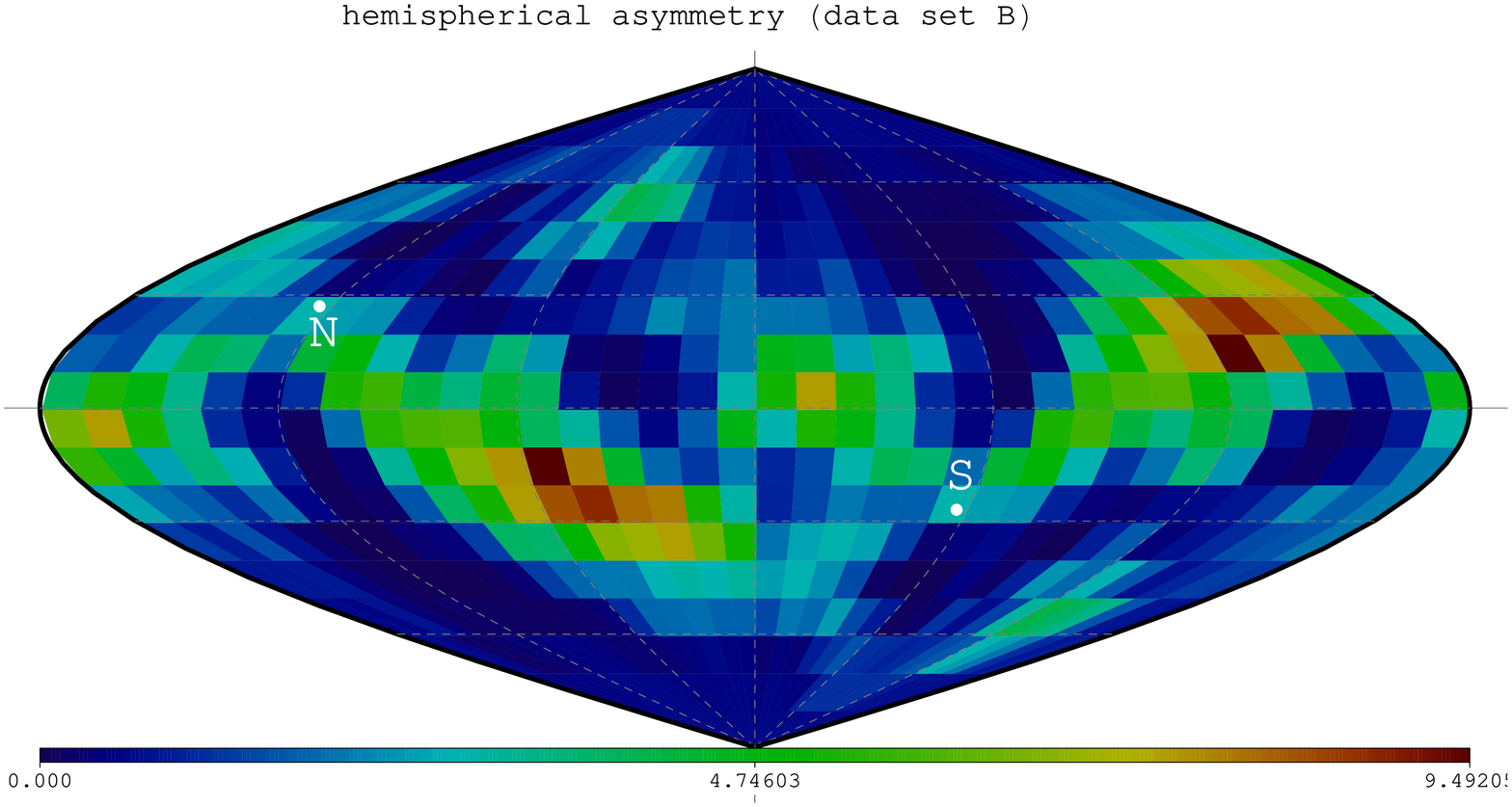} \caption{\label{fig8}
Hemispherical asymmetry in $\Delta\chi^2$ (see also figure
\ref{fig5}) in the context of the flat $\Lambda$CDM model. All SNe
with $A_V < 1.0$ at any redshift are included in the fits.}
\end{figure}

Finally, we search again for the maximally asymmetric pair of hemispheres.
In figure \ref{fig8} we show the hemispherical asymmetry $\Delta \chi^2$ in galactic coordinates.
For data set A, the location of the maxima is very similar to the location of maxima in figure \ref{fig5},
and we confirm our findings above.

In contrast, data set B shows a new structure, with a significant maximum close to $(l,b) \sim
(235^\circ, 15^\circ)$. This direction does not reflect any obvious large scale structure, but
as we average over half of the sky, the pointing is not expected to be precise. However, an
inspection of the asymmetry in the number of degrees of freedom (dof) of the hemispheres
shows that for data set B the maximum asymmetry in $\chi^2$ coincides with the maximal asymmetry in the number of dof. We conclude that the pattern for data set B is due to the geometry of the SN
sample and is fully consistent with an isotropic Hubble diagram. However, this finding does not rule
out the precence of the effect observed in data set A for two reasons: data set A contains significantly
more SNe and is dominated by nerby SNe ($\bar z = 0.27$), whereas set B is dominated by objects
at higher redshifts ($\bar z = 0.40$).

As above we run MCs to quantify the statistical significance of our findings. It turns out that the
pattern and amount of asymmetry in set B is not unexpected, still data set A is unexpected at the
$95\%$~C.L. (see table~\ref{LCDMgaleqHoqo}).

\subsubsection{Summary of the $\Lambda$CDM test}

We confirm the findings of the model-independent test, namely a significant
asymmetry in the SN calibration ($H_0/H_0^*$)  between North and South
equatorial hemispheres. On top of that asymmetries show up (especially in set B) that seem to be linked
to the geometry of the samples. We do not find significant evidence for large scale structure effects
at the largest scales ($180^\circ$), but as objects like the Shapely supercluster are significantly smaller,
our test is not well suited to identify such structures.

\section{Results}
\label{results}

\subsection{Is the Hubble diagram isotropic?}

We argue above that we find statistically significant deviations of
Hubble diagrams form isotropy. These are most significant in data sets
A and D, which contain most SNe. These asymmetries cannot be explained by
the peculiar motion of the observer, as our analysis is done in the CMB
rest-frame. If the asymmetry would be caused by peculiar motions of the
SNe hosts, we would expect that its significance decreases when
we include SNe at higher redshifts. This is not the case. As we go from the
model-independent test at small redshifts ($z<0.2$) to the model dependent
test for arbitrary redshift, the significance of the effect is actually
increased. On top of that, the magnitude of the observed
effect ($10 \%$ in distance scale) exceeds our expectations for the
effect from peculiar velocities ($v/(c\bar{z}) \sim 10^{-3}/\bar{z}$),
which in the most extreme case could explain effects at the
few per cent level in distance scale (or the SN calibration). However,
it is not excluded that large scale bulk motions contribute to the
effect, see e.g.~Cooray \& Caldwell (\cite{Cooray}) who argue that a
$5\%$ variation of $H$ might be possible in a low density bubble.
Thus the origin of the anisotropy is most likely either due to a systematic effect in SN search, observation or data analysis or a (large)
statistical fluke with a chance below 1:100.

Evidence that we can also exclude the possibility of a statistical
fluke comes from the fact that the hemispheres of maximal deviation
have poles close to the poles of the equatorial system. This suggests
that the origin of the observed anisotropy is, at least partly, due to
systematic errors. A possible candidate for a systematic error
would be inhomogeneous covering of the North and South sky, but
our MC studies reveal that the asymmetry in the number of the objects
cannot be held responsible. Exchanging the coordinates of the SNe
(which preserves the asymmetry in number of objects) typically produces
skies that are in agreement with the expectation of isotropy.  Another
possible explanation might be that telescopes or search strategies in the
North and South equatorial hemispheres have a systematic calibration off-set.

Yet another hint in favour of an unknown systematic effect is the pattern
observed in figures \ref{fig5} and \ref{fig8}. If the Hubble diagrams
would be isotropic, we would expect that there are several minima and maxima
distributed randomly on the sky. However we observe quite regular patterns in
the asymmetry maps. As the asymmetries in data set B are smaller than in data
sets A and D, it seems better suited for cosmological analysis. The magnitude
of its asymmetries could be consistent with large scale structure ($1\%$
effects), but are not statistically significant. However, the reason for the
consistency of data set B with isotropy might just be that it does
not contain enough SNe. Note that the fit values and their error bars of
set B are consistent with the fits of the asymmetric sets A and D.

Putting together all the evidence on the local flow direction in the
Universe, mainly based on the analysis of galaxy clusters, Hudson et
al.~(\cite{Hudson}) suggest a bulk flow of $225$ km/s toward
$(l,b)=(300^\circ, 10^\circ)$ at depth greater than $60 h^{-1}$ Mpc.
One would expect that this should be reflected by SNe. However, our
maximal asymmetric direction is significantly off-set from the large
scale galaxy cluster bulk flow. A closer inspection of figure \ref{fig5} for
sets A and D shows a modest, statistically insignificant, asymmetry
close to the suggested bulk flow direction.

\subsection{Is the local Hubble expansion rate different from the global one?}

We cannot answer that question, as we have no handle in our analysis to
the absolute calibration of SNe. If one could exclude that the calibration
asymmetries found in this work are due to systematic errors, this would be
a strong indication for a variation of the Hubble rate of the order of $10\%$.
If so, the tension between the WMAP 3yr measurment of $H_0$ and the
determination based on SNe by Sandage et al.~(\cite{Sandage:2006cv})
might be reconciled with each other.

The order of magnitude would also be consistent with recent claims by
Jha et al.~(\cite{Jha}) about the existence of a ``Hubble bubble'';
but more recent analysis (Wang \cite{Wang}; Conley et
al.~\cite{Conley}) does not confirm the claim. Conley et al.~(\cite{Conley})
show that the evidence for a ``Hubble bubble'' depends crucially on how SN
colours are modelled. As we think that our test points towards an equatioral
North/South systematic, we do not think that this work should be
regarded as a support of the idea of a local bubble. However, it might
be interesting to check if an equatorial systematic in SN colours could be
found.

\subsection{Is the Universe accelerating today?}

Based on our analysis one cannot answer this question with a straight
`yes'. The model-independent analysis is fully consistent with an
accelerated Universe, but the evidence in favour of acceleration is at
most at the $2 \sigma$ level. For the supposedly most accurate compilation
of local SNe from Jha et al.~\cite{Jha} (data set D), we find that $q_0 = 0$ is within
the $2\sigma$ contour (see our figure \ref{IFSHoqoTBRA}). The same is true for all four
analysed sets of data (including the SNLS data set, not shown in our figure
\ref{IFSHoqoTBRA}). What can be confirmed is that the Einstein-de Sitter model
($q_0 = 1/2$) is outside the $2\sigma$ region, and is
therefore disfavoured already by the model-independent test at small redshift.

If we assume that the flat $\Lambda$CDM is the full truth, then the SNe Ia
data can indeed provide convincing evidence that the Universe is accelerating.
Within that model deceleration ($q_0 \geq 0$) corresponds to
$\Omega_{\rm M} \geq 2/3$.  As can be seen from figure
\ref{OmH0TBRJ}, this case is excluded at high confidence. However,
we should keep in mind that neither the physics of `$\Lambda$', nor that
of `CDM' is understood at any depth.

\section{Conclusion}
\label{conclusion}

The purpose of the presented study is to develop tools for and to test the isotropy of Hubble
diagrams, which are at the very foundation of modern cosmology. Within our established
cosmological model we expect some small deviations from the isotropy, but with the
present day accuracy of SNe observations, we would not expect to be able to detect them
at high statistical significance (apart from our proper motion).
Nevertheless, we set out to apply the hemisphere tests to existing data in order to develop
the methodology and to test for systematic effects.

To our surprise we identified a statistically significant asymmetry, which is maximised close to the
orientation of the equatorial system. It seems to us that this calls for a thorough investigation of
possible systematic effects, which is beyond the scope of this work. Our analysis indicates that there is an off-set in the calibration of SNe between the equatorial hemispheres. Typically SN searches
are flux-limited (not red-shift limited) and thus one could imagine that the search in the North and in
the South selects samples with different dispersion and mean value in SN brightness.
Our findings would be consistent with a more complete search for nearby SN in the South compared to the North.
We think that this study shows, besides the interest in the large scale structure, that a large sky coverage of SN search missions is also important for the issue of systematic errors.

The Hubble diagram is currently the only direct mean to probe the
acceleration of the Universe. Most of our evidence for the present day
acceleration comes from indirect arguments and relies on a bunch of
untested assumptions. Our model-independent test fails to detect
acceleration of the Universe at high statistical significance.
It seems to us that it is too early to take
accelerated expansion of the Universe for granted, as the evidence
heavily relies on the {\em a priori} assumption of the $\Lambda$CDM
model.

\begin{table*}
\centering
\begin{tabular}{lcccc}

&dof & $\frac{\chi^2}{\rm dof}$ & $\frac{H_0}{H_0^*}$ & $q_0$ \\
\hline
{\bf data set A:} 253 SNe, $z \in [0.002,1.755]$ \\
$A_V \leq 1$, $\sigma_v = 345$ km/s, $z \leq 0.2$
                      & 137 & 1.27 & $1.02 \pm 0.02$ & $-0.78 \pm 0.90$ \\
$A_V \leq 0.5$        & 116 & 1.30 & $1.02 \pm 0.02$ & $-0.56 \pm 0.94$ \\
$\sigma_v = 230$ km/s & 137 & 1.82 & $1.03 \pm 0.02$ & $-0.97 \pm 0.85$ \\
$\sigma_v = 460$ km/s & 137 & 0.97 & $1.02 \pm 0.03$ & $-0.68 \pm 0.96$ \\
$\sigma_v = 690$ km/s & 137 & 0.67 & $1.02 \pm 0.03$ & $-0.61 \pm 1.06$ \\
%$\sigma_v = 1150$ km/s & 137 & 0.40 & $1.02 \pm 0.04$ & $-0.61 \pm 1.26$ \\
$0.02 < z \leq 0.2$   &  73 & 1.20 & $1.01 \pm 0.03$ & $-0.31 \pm 1.08$ \\
$z \leq 0.1$          & 128 & 1.33 & $1.02 \pm 0.03$ & $-0.52 \pm 0.52$ \\
$0.01 < z \leq 0.1$   &  97 & 1.11 & $1.00 \pm 0.03$ & $ \ \ 0.28 \pm 1.25$ \\
$z \leq 0.3$          & 142 & 1.23 & $1.02 \pm 0.02$ & $-0.65 \pm 0.73$ \\
\hline
{\bf data set B:} 186 SNe, $z\in [0.010, 1.755]$ \\
$A_V \leq 1$, $\sigma_v = 345$ km/s, $z \leq 0.2$
                      & 75 & 0.84 & $1.01 \pm 0.03$ & $-1.42 \pm 1.23$ \\
$A_V \leq 0.5$        & 66 & 0.79 & $1.00 \pm 0.04$ & $-1.36 \pm 1.79$ \\
$\sigma_v = 230$ km/s & 75 & 0.92 & $1.01 \pm 0.03$ & $-1.43 \pm 1.20$ \\
$\sigma_v = 460$ km/s & 75 & 0.75 & $1.01 \pm 0.03$ & $-1.40 \pm 1.25$ \\
$\sigma_v = 690$ km/s & 75 & 0.60 & $1.01 \pm 0.04$ & $-1.35 \pm 1.33$ \\
%$\sigma_v = 1150$ km/s & 75 & 0.39 & $1.01 \pm 0.05$ & $-1.25 \pm 1.47$ \\
$0.02 < z \leq 0.2$   & 50 & 0.95 & $1.01 \pm 0.04$ & $-1.27 \pm 1.36$ \\
$z \leq 0.1$          & 70 & 0.84 & $1.02 \pm 0.05$ & $-2.16 \pm 2.21$ \\
$z \leq 0.3$          & 80 & 0.81 & $1.00 \pm 0.03$ & $-0.78 \pm 0.89$ \\
\hline
{\bf data set C:} 117 SNe, $z\in [0.015, 1.01]$ \\
$\sigma_v = 345$ km/s, $\sigma_{int} = 0.03$, $z \leq 0.2$
                      & 42 & 0.84 & $1.07 \pm 0.04$ & $-0.57 \pm 1.63$ \\
$\sigma_v = 230$ km/s & 42 & 0.92 & $1.08 \pm 0.04$ & $-0.57 \pm 1.59$ \\
$\sigma_v = 460$ km/s & 42 & 0.76 & $1.08 \pm 0.05$ & $-0.58 \pm 1.71$ \\
$\sigma_v = 690$ km/s & 42 & 0.60 & $1.08 \pm 0.06$ & $-0.63 \pm 1.88$ \\
%$\sigma_v = 1150$ km/s & 42 & 0.38 & $1.08 \pm 0.07$ & $-0.75 \pm 2.25$ \\
$0.02 < z \leq 0.2$   & 30 & 1.07 & $1.07 \pm 0.05$ & $-0.28 \pm 1.83$ \\
$z \leq 0.1$          & 40 & 0.87 & $1.06 \pm 0.05$ & $ \ \ 0.16 \pm 2.15$ \\
$z \leq 0.3$          & 46 & 0.80 & $1.08 \pm 0.03$ & $-0.56 \pm 0.58$ \\
$\sigma_{int}=0$      & 42 & 3.73 & $1.09 \pm 0.03$ & $-0.82 \pm 0.75$ \\
$\sigma_{int}=0.02$   & 42 & 1.40 & $1.08 \pm 0.04$ & $-0.62 \pm 1.27$ \\
\hline
{\bf data set D:} 131 SNe, $z \in [0.002, 0.124]$ \\
$A_V \leq 1$, $\sigma_v = 345$ km/s, $\sigma_{int} = 0.016$, $z \leq
0.2$
                      & 117 & 1.37 & $1.01 \pm 0.03$ & $-1.39 \pm 1.35$ \\
$A_V \leq 0.5$        &  99 & 1.43 & $1.00 \pm 0.02$ & $-0.86 \pm 1.36$ \\
$\sigma_v = 230$ km/s & 117 & 2.14 & $1.02 \pm 0.02$ & $-1.68 \pm 1.26$ \\
$\sigma_v = 460$ km/s & 117 & 0.98 & $1.01 \pm 0.03$ & $-1.16 \pm 1.45$ \\
$\sigma_v = 690$ km/s & 117 & 0.59 & $1.00 \pm 0.04$ & $-0.86 \pm 1.60$ \\
%$\sigma_v = 1150$ km/s & 117 & 0.29 & $1.00 \pm 0.05$ & $-0.60 \pm 1.94$ \\
$0.02 < z \leq 0.2$   & \ 50 & 1.24 & $0.97 \pm 0.04$ & $\ 0.11 \pm 1.59$ \\
$z \leq 0.1$          & 115 & 1.39 & $1.02 \pm 0.03$ & $-1.49 \pm 1.71$ \\
$0.01 < z \leq 0.1$   & 84 & 1.08 & $1.00 \pm 0.03$ & $-0.81 \pm 1.78$ \\
$\sigma_{int}=0$      & 117 & 1.64 & $1.01 \pm 0.02$ & $-1.30 \pm 1.20$ \\
$\sigma_{int}=0.03$   & 117 & 1.05 & $1.02 \pm 0.03$ & $-1.51 \pm 1.66$ \\
\hline
\end{tabular}
\caption{\label{TFSHoqo}Robustness of the full-sky fit of the
calibration $H_0/H_0^*$ and the deceleration parameter $q_0$ at
small redshifts. We compare the number of degrees of
freedom (dof), $\chi^2$/dof, and the best fit cosmological
parameters for the four data sets described in the text for various
assumptions on
acceptable light extinction $A_V$, peculiar velocity dispersion $\sigma_v$, intrinsic dispersion
$\sigma_{\rm int}$, as well as redshift interval included in the fit. Our analysis with $A_V < 1$
includes all SNe without information on $A_V$, but we exclude those when investigating $A_V < 0.5$.}
\end{table*}

\begin{table*}
\centering
\begin{tabular}{lccccccc}
& dof & $\frac{\chi^2}{\rm dof}$ & $\frac{H_0}{H_0^*}$ & $q_0$ & MC$_{\chi^2}$  & MC$_{\frac{H_0}{H_0^*}}$  & MC$_{q_0}$  \\
&     &                          &                     &       &
($\Delta\chi^2$) & ($\Delta\frac{H_0}{H_0^*}$)& ($\Delta q_0$)\\
\hline
{\bf data set A} \\
$A_V \leq 1$, $\sigma_v = 345$ km/s, $z \leq 0.2$
                      & 137 & 1.27 & $1.02 \pm 0.02$ & $-0.78 \pm 0.90$ &&& \\
Galactic hemispheres & & & & & & & \\
North: $(l,b)=(0^{\circ},90^{\circ})$   & 72 & 1.34 &$ 1.05 \pm 0.03 $&$ -1.92 \pm 1.53$& 32.6$\%$ & 21.2$\%$  & 14.2$\%$  \\
South: $(l,b)=(0^{\circ},-90^{\circ})$  & 63 & 1.18 & $1.00 \pm 0.03 $&$ \quad0.03 \pm 1.14$&(4.75)&(0.05)&(1.95)  \\
Equatorial hemispheres & & & & & & & \\
North: $(l,b)=(123^{\circ},27^{\circ})$ & 70 & 1.35 &$ 1.01 \pm 0.03$&$ -1.31 \pm 1.63$& 0.8$\%$  & 4.6$\%$  & 81.6$\%$  \\
South: $(l,b)=(303^{\circ},-27^{\circ})$& 65 & 1.07 &$ 1.07 \pm 0.04$&$ -1.57 \pm 1.38$&(21.71)&(0.06)&(0.26) \\
Hemispheres max. Asymmetry in $\chi^2$:& & & &  &&& \\
Pole: $(l,b)=(110^{\circ},24^{\circ})$ & 65 & 0.89 &$ 0.99 \pm 0.03 $&$ -1.02 \pm 1.75$  &&&\\
Pole: $(l,b)=(290^{\circ},-24^{\circ})$& 70 & 1.38 &$ 1.09 \pm 0.04 $&$ -2.12 \pm 1.29$  &&&\\
\hline
{\bf data set B} \\
$A_V \leq 1$, $\sigma_v = 345$ km/s, $z \leq 0.2$
                      & 75 & 0.84 & $1.01 \pm 0.03$ & $-1.42 \pm 1.23$ &&&\\
Galactic hemispheres & & & & &&&\\
North: $(l,b)=(0^{\circ},90^{\circ})$  & 41 & 0.80 &$ 1.02 \pm 0.04$ & $-1.75 \pm 1.69$& 81.4$\%$  & 56.2$\%$  & 59.8$\%$ \\
South: $(l,b)=(0^{\circ},-90^{\circ})$ & 32 &0.93& $1.00 \pm 0.05 $& $-0.99 \pm 1.80$ &(0.49)&(0.02)&(0.76)\\
Equatorial hemispheres & & & & &&&\\
North: $(l,b)=(123^{\circ},27^{\circ})$ & 41 & 0.80 &$ 1.01 \pm 0.04 $&$ -2.02 \pm 1.79$ & 15.2$\%$ & 54.2$\%$ & 61.6$\%$  \\
South: $(l,b)=(303^{\circ},-27^{\circ})$& 32 & 0.86 &$ 1.03 \pm 0.06$&$ -1.32 \pm 1.85$ &(5.36)&(0.02)&(0.70)\\
Hemispheres max. Asymmetry in $\chi^2$:& & & & &&& \\
Pole: $(l,b)=(62^{\circ},11^{\circ})$ & 33 & 0.57 &$ 0.97 \pm 0.05 $&$ -1.26 \pm 2.62$   &&& \\
Pole: $(l,b)=(242^{\circ},-11^{\circ})$& 40 & 0.93 &$ 1.08 \pm 0.05 $&$ -2.23 \pm 1.56$  &&& \\
\hline
{\bf data set D} \\
$A_V \leq 1$, $\sigma_v = 345$ km/s, $\sigma_{int}=0.016$ z$ \leq
0.2$
                      & 117 & 1.37 & $1.01 \pm 0.03$ & $-1.39 \pm 1.35$ &&&\\
Galactic hemispheres & & & & &&&\\
North: $(l,b)=(0^{\circ},90^{\circ})$  & 69 & 1.49 &$ 1.02 \pm 0.03 $& $ -1.85\pm 2.03 $&82.8$\%$& 89.4$\%$& 73.6$\%$ \\
South: $(l,b)=(0^{\circ},-90^{\circ})$ & 46 & 1.25 &$ 1.01 \pm 0.04 $& $ -1.07\pm 1.84$ &(0.82)&(0.004)&(0.77)\\
Equatorial hemispheres & & & & &&&\\
North: $(l,b)=(123^{\circ},27^{\circ})$ & 64 & 1.71 &$ 0.99 \pm 0.03$&$ -0.85\pm 2.12$ & 7.8$\%$& 5.0$\%$& 31.8$\%$\\
South: $(l,b)=(303^{\circ},-27^{\circ})$& 51 & 0.87 &$ 1.06 \pm 0.04$&$ -2.92\pm 1.98$ &(10.84)&(0.075)&(2.07)\\
Hemispheres max. Asymmetry in $\chi^2$:& & & & &&& \\
Pole: $(l,b)=(99^{\circ},24^{\circ})$ & 61 & 1.15 &$ 0.96 \pm 0.03 $&$ 0.96 \pm 1.85$   &&& \\
Pole: $(l,b)=(279^{\circ},-24^{\circ})$& 54 & 1.31 &$ 1.10 \pm 0.04 $&$ -4.73 \pm 2.14$  &&& \\
\hline
&&&& \\
\end{tabular}
\caption{\label{TgaleqHoqo}Hemisphere fit of $H_0/H_0^*$ and $q_0$
at small redshifts. We compare the number of degrees of freedom
(dof), $\chi^2$/dof, and the best fit cosmological parameters for
the data sets A, B and D for the galactic, equatorial and maximal
asymmetry hemispheres. In addition we show for the galactic and
equatorial hemispheres the percentage of MC simulations with a
larger deviation in $\chi^2$, $H_0/H_0^*$ and $q_0$. In
brackets we provide the respective differences $\Delta \chi^2$,
$\Delta \frac{H_0}{H_0^*}$ and $\Delta q_0$.}
\end{table*}

\begin{table*}
\centering
\begin{tabular}{lccccccc}
& dof & $\frac{\chi^2}{\rm dof}$ & $\frac{H_0}{H_0^*}$ & $q_0$ &
MC$_{\chi^2}$ & MC$_{\frac{H_0}{H_0^*}}$ & MC$_{q_0}$  \\
&  &  &  &  &  ($\Delta\chi^2$)&
($\Delta\frac{H_0}{H_0^*}$) &($\Delta q_0$) \\
\hline
{\bf data set A} \\
$A_V \leq 1$, $\sigma_v = 345$ km/s, $z \leq 0.2$
                      & 137 & 1.27 & $1.02 \pm 0.02$ & $-0.78 \pm 0.90$ &&& \\
Hemispheres max. Asymmetry in $\chi^2$:& & & &&&&\\
Pole: $(l,b)=(60^{\circ},10^{\circ})  $& 53 & 0.84 &$ 0.97 \pm 0.03 $&$ -0.11 \pm 1.45$ & 0.2$\%$ & 4.0$\%$ & 48.6$\%$  \\
Pole: $(l,b)=(240^{\circ},-10^{\circ})$& 82 & 1.33 &$ 1.08 \pm 0.03 $&$ -2.05 \pm 1.28$ & (37.93)   & (0.11) & (1.95)  \\
\hline
{\bf data set B} \\
$A_V \leq 1$, $\sigma_v = 345$ km/s, $z \leq 0.2$
                      & 75 & 0.84 & $1.01 \pm 0.03$ & $-1.42 \pm 1.23$ &&&\\
Hemispheres max. Asymmetry in $\chi^2$:& & & &&&&\\
Pole: $(l,b)=(60^{\circ},10^{\circ})  $& 34 & 0.57 &$ 0.98 \pm 0.05 $&$ -1.34 \pm 2.63$ & 38.6$\%$ & 50.2$\%$ & 71.4$\%$  \\
Pole: $(l,b)=(240^{\circ},-10^{\circ})$& 39 & 0.96 &$ 1.06 \pm 0.05 $&$ -2.20 \pm 1.56$ & (11.98)   & (0.08 & (0.87) ) \\
\hline
{\bf data set D} \\
$A_V \leq 1$, $\sigma_v = 345$ km/s, $\sigma_{int}=0.016$ z$ \leq
0.2$
                      & 117 & 1.37 & $1.01 \pm 0.03$ & $-1.39 \pm 1.35$ &&&\\
Hemispheres max. Asymmetry in $\chi^2$:& & & &&&&\\
Pole: $(l,b)=(75^{\circ},20^{\circ})  $& 53 & 1.14 &$ 0.95 \pm 0.03 $&$ 1.45 \pm 1.88$ & $<$0.2$\%$  & 0.2$\%$ & 1.8$\%$ \\
Pole: $(l,b)=(255^{\circ},-20^{\circ})$& 62 & 1.36 &$ 1.08 \pm 0.04 $&$ -4.17 \pm 1.98$ & (32.33)   & (0.13)  & (5.62) \\
\hline
&&&&&&& \\
\end{tabular}
\caption{\label{tab4}Hemispheres of maximal asymmetry from a $5^\circ$
search (in contrast to the more accurate $1^\circ$ search in table
\ref{TgaleqHoqo}). The purpose of the sparse search is to run MC
simulations. The meaning of the columns is explained in the caption of table \ref{TgaleqHoqo}.}
\end{table*}

\begin{table*}
\centering
\begin{tabular}{lcccc}
 & dof & $\frac{\chi^2}{\rm dof}$ & $\frac{H_0}{H_0^*}$ & $\Omega_{\rm M}$ \\
\hline
{\bf data set A:} 253 SNe, $z \in [0.002,1.755]$\\
$A_V \leq 1$, $\sigma_v = 345$ km/s, $z$ arbitrary & 242 &1.31 & $  1.02\pm0.02 $&$  0.31\pm0.06 $ \\
$\sigma_v = 230$ km/s & 242 & 1.62 &$ 1.02 \pm 0.02$ & $ 0.31 \pm 0.06$ \\
$\sigma_v = 460$ km/s & 242 & 1.14 &$ 1.01 \pm 0.02$ & $ 0.34 \pm 0.06 $\\
$\sigma_v = 690$ km/s & 242 & 0.96 &$ 1.01 \pm 0.02$ & $ 0.34 \pm 0.07 $\\
%$\sigma_v = 1150$ km/s & 242 & 0.80 &$ 1.01 \pm 0.02$ & $ 0.34 \pm 0.08$ \\
$0.0 < z \leq 0.2$    & 137 & 1.27 &$ 1.03 \pm 0.03$ & $ 0.04^{ +0.66}_{ -0.04} $\\
$0.01<z$              & 211 & 1.20 &$ 1.01 \pm 0.02$ & $ 0.34 \pm 0.07 $\\
$A_V \leq 0.5$        & 219 & 1.26 &$ 1.01 \pm 0.02$ & $ 0.32 \pm 0.07 $\\
\hline
{\bf data set B:} 186 SNe, $z\in [0.010, 1.755]$\\
$A_V \leq 1$, $\sigma_v = 345$ km/s, $z$ arbitrary & 180 & 1.11 & $ 0.98 \pm 0.02$&$ 0.32 \pm0.06 $ \\
$\sigma_v = 230$ km/s & 180 & 1.14 & $0.98\pm 0.02$ & $ 0.32\pm 0.06 $ \\
$\sigma_v = 460$ km/s & 180 & 1.07 & $ 0.98\pm 0.02$ &$ 0.32 \pm 0.06 $\\
$\sigma_v = 690$ km/s & 180 & 1.00 & $ 0.98\pm 0.02$& $0.32 \pm 0.06 $\\
%$\sigma_v = 1150$ km/s & 180 & 0.91 & $ 0.97\pm 0.03 $&$ 0.34 \pm 0.08$ \\
$0.0 < z \leq 0.2$      & \ 75 & 0.84 &$ 1.00\pm 0.02 $&$\leq$ 0.96 ($2\sigma$)\\
$A_V \leq 0.5$           & 159 & 0.94 &$ 0.97 \pm 0.02 $&$ 0.32 \pm 0.06 $\\
\hline
{\bf data set C:} 117 SNe, $z\in [0.015, 1.01]$\\
$A_V \leq 1$, $\sigma_v = 345$ km/s, $\sigma_{int}=0.03$, $z$ arbitrary & 115 & 1.02 & $ 1.08 \pm 0.03$&$ 0.25 \pm 0.07$ \\
$\sigma_v = 230$ km/s & 115 & 1.05 & $1.08\pm 0.05$ & $ 0.23\pm 0.07 $ \\
$\sigma_v = 460$ km/s & 115 & 0.99 & $ 1.08\pm 0.02$ &$ 0.23 \pm 0.06 $\\
$\sigma_v = 690$ km/s & 115 & 0.93 & $ 1.08\pm 0.03 $& $0.23 \pm 0.06 $\\
%$\sigma_v = 1150$ km/s & 115 & 0.85 & $ 1.07\pm 0.03 $&$ 0.26 \pm 0.08 $ \\
$0.0 < z \leq 0.2$      & \ 42 & 0.84 &$ 1.08 \pm 0.04 $&$ 0.20^{ +1.48}_{ -0.20} $\\
$\sigma_{int}=0 $     & 115 & 8.60 & $ 1.08 \pm 0.01$&$ 0.27 \pm 0.03$ \\
$\sigma_{int}=0.02  $ & 115 & 1.68 & $ 1.08 \pm 0.05$&$ 0.24 \pm 0.02$ \\
\hline
\end{tabular}
\caption{\label{tabLCDMfull}Robustness of the full-sky fit to the
flat $\Lambda$CDM model at arbitrary redshift. The fit parameters
are the calibration $H_0/H_0^*$ and the dimensionless matter density
$\Omega_{\rm M}$. We compare the number of degrees of freedom (dof),
$\chi^2$/dof, and the best fit cosmological parameters for data sets
A, B, and C for various assumptions on the acceptable light
extinction $A_V$, peculiar velocity dispersion $\sigma_v$, intrinsic dispersion $\sigma_{\rm int}$
and redshift interval included in the fit.}
\end{table*}

\begin{table*}
 \centering
\begin{tabular}{lccccccc}
& dof & $\frac{\chi^2}{\rm dof}$ & $\frac{H_0}{H_0^*}$ & $\Omega_{\rm M}$
& MC$_{\chi^2}$ & MC$_{\frac{H_0}{H_0^*}}$ & MC$_{\Omega_{\rm M}}$ \\
&  &  &  &&  ($\Delta\chi^2$) &
($\Delta\frac{H_0}{H_0^*}$) & ($\Delta \Omega_{\rm M}$) \\
\hline
{\bf data set A} \\
$A_V \leq 1$, $\sigma_v = 345$ km/s, $z$ arbitrary
                      & 242 &1.31 & $  1.02\pm0.02 $&$  0.31\pm0.06 $&&& \\
Galactic hemispheres & & & & & & & \\
North: $(l,b)=(0^{\circ},90^{\circ})$   & 136 & 1.36 &$ 1.02 \pm
0.02 $&$
0.24 \pm 0.07$& 6.47$\%$ & 12.80$\%$  & 5.00$\%$  \\
South: $(l,b)=(0^{\circ},-90^{\circ})$  & 104 & 1.21 &$ 1.00 \pm
0.02 $&$
0.44 \pm 0.12$&(9.11)&(0.02)& (0.20) \\
Equatorial hemispheres & & & & & & & \\
North: $(l,b)=(123^{\circ},27^{\circ})$ & 136 & 1.32 &$ 0.99 \pm
0.02$&$
0.34 \pm 0.08$& 0.2$\%$  & 0.6$\%$  & 64.6$\%$  \\
South: $(l,b)=(303^{\circ},-27^{\circ})$& 104 & 1.18 &$ 1.04 \pm
0.02$&$
0.35 \pm 0.11$&(27.15)&(0.05)& (0.01)\\
Hemispheres max. Asymmetry in $\chi^2$:& & & &&&&\\
Pole: $(l,b)=(120^{\circ},25^{\circ}) $& 133 & 1.25 &$ 0.98 \pm 0.02
$&$
0.36 \pm 0.08$ & 4.6$\%$  & 12.2$\%$ & 93.8$\%$ \\
Pole: $(l,b)=(300^{\circ},-25^{\circ})$& 107 & 1.24 &$ 1.04 \pm 0.02
$&$
0.40 \pm 0.10$ & (35.08)    & (0.06) & (0.04) \\
\hline
{\bf data set B} \\
$A_V \leq 1$, $\sigma_v = 345$ km/s, $z \leq 0.2$
                      & 75 & 0.84 & $1.01 \pm 0.03$ & $-1.42 \pm 1.23$ &&&\\
Galactic hemispheres & & & & &&&\\
North: $(l,b)=(0^{\circ},90^{\circ})$  & 106 & 1.00 &$ 1.00 \pm
0.03$ & $
0.28 \pm 0.08$& 38.21$\%$  & 9.17$\%$  & 37.98$\%$ \\
South: $(l,b)=(0^{\circ},-90^{\circ})$ & 72 & 1.28 & $ 0.97\pm 0.03
$& $
0.34\pm 0.09$ &(1.89)&(0.03)&(0.06)\\
Equatorial hemispheres & & & & &&&\\
North: $(l,b)=(123^{\circ},27^{\circ})$ & 108 & 1.12 &$ 0.97 \pm
0.03 $&$
0.31 \pm 0.08$ & 7.8$\%$ & 4.2$\%$ & 94.2$\%$  \\
South: $(l,b)=(303^{\circ},-27^{\circ})$& 70 & 1.08 &$ 1.00 \pm
0.03$&$ 0.32
\pm 0.09$ &(5.86)&(0.03)&(0.01)\\
Hemispheres max. Asymmetry in $\chi^2$:& & & &&&&\\
Pole: $(l,b)=(235^{\circ},15^{\circ})$& 130 & 1.03 &$ 1.00 \pm 0.03
$&$ 0.30
\pm  0.07 $& 62.0$\%$  & 73.3$\%$ & 27.7$\%$ \\
Pole: $(l,b)=(55^{\circ},-15^{\circ})$&  48 & 1.13 &$ 0.98 \pm 0.03
$&$ 0.12 \pm  0.11 $ & (13.84)    & (0.02) & (0.18) \\
\hline
&&&& \\
\end{tabular}
\caption{\label{LCDMgaleqHoqo}Comparison of galactic, equatorial and
maximal asymmetry hemispheres for the flat $\Lambda$CDM model. We
use all SNe with $A_V < 1.0$ from data sets A and B. The meaning of the
columns is explained in the caption of table \ref{TgaleqHoqo}.}
\end{table*}

\begin{acknowledgements}
We thank Camille Bonvin, Ruth Durrer, Chris Gordon, Steen Hannestad,
Stefan Hofmann, Michael Hudson, Ariel Goobar, Martin Kunz,
Bruno Leibundgut and Anze Slozar for discussions and comments.
\end{acknowledgements}

\end{document}